\documentclass[aoas, preprint]{imsart}

\RequirePackage{amsthm,amsmath,amsfonts,amssymb}
\RequirePackage[authoryear]{natbib}
\RequirePackage[colorlinks,citecolor=blue,urlcolor=blue]{hyperref}
\RequirePackage{graphicx}
\usepackage{subcaption, cleveref}
\usepackage{algorithm, algpseudocode}
\usepackage{booktabs, rotating, multirow}
\usepackage{url}
\usepackage{anyfontsize}
\usepackage{placeins}

\startlocaldefs

\newcommand{\bth}{\boldsymbol{\theta}} \newcommand{\bx}{\mathbf{x}}
\newcommand{\bX}{\mathbf{X}} \DeclareMathOperator*{\argmin}{arg\,min}
\newcommand{\bz}{\mathbf{z}}
\newcommand{\norm}[1]{\left\lVert#1\right\rVert}
\newcommand{\bbeta}{\boldsymbol{\beta}} \newcommand{\bI}{\mathbf{I}}
\newcommand{\bJ}{\mathbf{J}} 
\newcommand{\bY}{\mathbf{Y}} 

\algrenewcommand\algorithmicrequire{\textbf{Input:}}
\algrenewcommand\algorithmicensure{\textbf{Output:}}

\endlocaldefs

\begin{document}

\begin{frontmatter}
\title{Neural Posterior Estimation for Stochastic Epidemic Modeling}
\runtitle{Neural Posterior Estimation for Stochastic Epidemic Modeling}

\begin{aug}

\author[A]{\fnms{Prayag}~\snm{Chatha}\ead[label=e1]{pchatha@umich.edu}}
\author[B]{\fnms{Fan}~\snm{Bu}\ead[label=e2]{fbu@umich.edu}}
\author[A]{\fnms{Jeffrey}~\snm{Regier}\ead[label=e3]{regier@umich.edu}}
\author[C]{\fnms{Evan}~\snm{Snitkin}\ead[label=e4]{esnitkin@umich.edu}}
\author[D]{\fnms{Jon}~\snm{Zelner}\ead[label=e5] {jzelner@umich.edu}}
\address[A]{Department of Statistics, University of Michigan \printead[presep={
,\ }]{e1,e3}}
\address[B]{Department of Biostatistics, University of Michigan \printead[presep={,\ }]{e2}}
\address[C]{Department of Microbiology and Immunology, University of Michigan \printead[presep={,\ }]{e4}}
\address[D]{Department of Epidemiology \& Center for Social Epidemiology and
Population Health, University of Michigan \printead[presep={,\ }]{e5}}
\end{aug}

\begin{abstract}
Stochastic infectious disease models capture uncertainty in public health
outcomes and have become increasingly popular in epidemiological practice.
However, it is hard to calibrate realistic stochastic models to data due to the
challenges of likelihood-based inference of unknown parameters. Stochastic
epidemic models are nonlinear dynamical systems that may feature massive latent
state spaces, resulting in computationally intractable likelihood densities. We
develop an approach to calibrating large-scale epidemiological models using
Neural Posterior Estimation, an emergent deep learning technique for
simulation-based inference. In NPE, a neural network trained on simulated data
learns to ``invert'' a stochastic simulator, returning a parametric approximation
to the posterior distribution. Motivated by the problem of understanding
transmission of carbapenem-resistant Klebsiella pneumoniae (CRKP), a major
healthcare-associated infection, we propose a stochastic, discrete-time
Susceptible Infected model. Through a realistic simulation experiment, we show
that NPE produces accurate posterior estimates of unknown infection rates at a
computational discount compared to Approximate Bayesian Computation. In an
empirical study of CRKP transmission in a Chicago-area hospital, we use NPE to
analyze spatial heterogeneity in patient-to-patient transmission risk.

\end{abstract}


\end{frontmatter}


\section{Introduction}

Complex, mechanistic transmission models are a critical tool for the practice of
infectious disease modeling and are increasingly used to make public health
decisions in real time (\cite{zelner2022rapid}). Epidemics are dynamical systems
that feature an interplay of both deterministic forces and random events
(\cite{Wood2010Statistical, hilborn2013ecological}). As computational resources
have become more available, stochastic models that generate a probability
distribution of epidemics have become prominent in epidemiology
(\cite{daley2001epidemic, britton2010stochastic}). Stochastic models reflect the
randomness of health outcomes in terms of both observational noise and the
aleatoric nature of disease transmission through a population (\cite{
he2010plug}), making them useful for informing public health policy under
conditions of uncertainty.

Parameters governing stochastic epidemic models are often inferred from
observational data. Bayesian inference offers a logical framework for estimating
unknown epidemiological parameters with uncertainty quantification and
calibrating large models to a given dataset (\cite{dunson2001commentary}).
However, the difficulty of likelihood-based parameter estimation, the mainstay
of classical statistical inference, has limited the design and feasibility of
stochastic epidemic models. Realistic transmission models are nonlinear
dynamical systems with a potentially large state space of latent variables (often
unobserved events such as infection onset times) that result in
intractable likelihood densities, thwarting conventional statistical tools
(\cite{breto2009time, endo2019introduction}). In particular, individual-level
models of infectious disease have likelihoods that may increase in exponential
complexity with respect to the size of the population
(\cite{cauchemez2011methods}). Recovering complete likelihoods involves
computationally intensive sampling methods such as data-augmented Markov chain
Monte Carlo (MCMC) that are tailored to a single model (\cite{bu2022likelihood,
fintzi2022linear}). When latent variables exhibit high serial correlation, as is
often the case for temporal models, mixing of chains can be prohibitively slow
(\cite{endo2019introduction}).

Forward sampling from stochastic dynamical systems is inexpensive in comparison
to evaluating their likelihoods. This property motivates a wide variety of
statistical methods that treat models of this type as black-box simulators.
Prominent among these are Iterated Filtering (\cite{ionides2006inference,
Ionides2011Iterated}), which uses sequential Monte Carlo to perform approximate
maximum likelihood (i.e. frequentist) inference for partially observed Markov
processes, and Approximate Bayesian Computation (ABC,
\cite{beaumont2010approximate, toni2010simulation}). ABC is an established tool
for fitting complex epidemiological models (\cite{McKinley2018Approximate,
Minter2019Approximate}), but can suffer from poor computational efficiency in
high dimensional settings and may be reliant on expert-designed summary
statistics (\cite{cranmer2020frontier, lueckmann2021benchmarking}). 

Neural Posterior Estimation (NPE) is a novel technique for approximate Bayesian
inference in which a neural network trained on simulated data predicts plausible
model parameters, effectively learning how to ``invert'' a forward simulation
model (\cite{papamakarios2016fast}). NPE can be understood as automatically
learning important features (i.e. summary statistics) from raw data in order
perform posterior inference without evaluation of an intractable likelihood. NPE
does not require extensive tailoring to a given simulator. Once trained, NPE can
draw posterior samples---conditioning on any dataset---cheaply and in parallel,
in contrast to Monte Carlo methods. NPE belongs to an emerging family of
simulation-based, deep learning-powered inference methods (e.g.
\cite{papamakarios2019neural}) that scale efficiently to high-dimensional data
(\cite{cranmer2020frontier}). These algorithms have been used for statistical
inference in a wide range of scientific domains, including neuroscience
(\cite{lueckmann2017flexible}), particle physics (\cite{baydin2019efficient}),
and astronomy (\cite{liu2023variational, dax2021real, vasist2023neural}).

In this work, we apply NPE to the general problem of calibrating complex
stochastic epidemiological models to observed data, a task for which established
statistical tools are often inadequate.\footnote{NPE and related neural methods
have been applied previously to deterministic epidemiological and ecological
models with observational noise; see
\cite{papamakarios2016fast,lueckmann2021benchmarking} and, for a notable
real-world study, \cite{radev2021outbreakflow}. However, in this deterministic setting,
nonlinear least squares or other standard optimization techniques are typically sufficient for parameter estimation
(\cite{chowell2017fitting}).} Our specific application concerns the analysis of
spatially heterogeneous transmission risks in an intervention study of CRKP, a
widespread and often lethal healthcare-associated infection (HAI), in a
long-term acute care hospital (LTACH). We find that NPE shows promise for
driving an efficient yet accurate simulation-based methodology for infectious
disease modeling, although this approach requires careful criticism of
inferential findings. 

The remainder of our article is organized as follows: in Section~\ref{sec:crkp},
we introduce the CRKP dataset and modeling difficulties motivating our
investigation. Next, we explicate the NPE algorithm in the context of Bayesian
simulation-based inference (Section~\ref{sec:sbi}). In
Section~\ref{sec:si-model}, we present a heterogeneous, discrete-time stochastic
susceptible-infected (SI) model suitable for healthcare-associated infections
such as CRKP. Through a realistic simulation experiment, we demonstrate the
scalability advantages of NPE over comparable calibration methods. Then, in
Section~\ref{sec:application}, we fit our stochastic model to the CRKP data in
order to analyze the spatial heterogeneity of transmission risks. We conclude
with a discussion of methodology and directions for future work
(Section~\ref{sec:disc}) and a statement of our work's significance to research
practice in statistical epidemiology (Section~\ref{sec:signif}). 

\section{Healthcare-associated Infection Study}\label{sec:crkp}
\subsection{Background Motivation}

Carbapenem-resistant Klebsiella pneumoniae (CRKP) is an enteric (originating in
the digestive system) bacterium  that first emerged in 2001 in American
hospitals. While CRKP colonization is generally asymptomatic, it may erupt into
an invasive infection (e.g. sepsis) if it enters internal organs. CRKP spreads
readily via skin contact, often indirectly via health care workers. Hospital
patients who are administered indwelling medical devices face elevated risk of
life-threatening CRKP infection (\cite{lledo2009guidance}). Accordingly, LTACHs
are major sites of CRKP incidence and mortality. As its name would suggest, CRKP
resists standard antibiotic treatment, so understanding its transmission is crucial for
developing effective mitigation strategies (\cite{han2019whole}).

Our data comes from an intervention study conducted in a Chicago-area LTACH from
2012 to 2013 that attempted to reduce CRKP transmission
(\cite{hayden2015prevention}). All patients were administered daily antiseptic
baths, those infected with CRKP were isolated in ward cohorts, and healthcare
workers received additional training in sanitation protocol and monitoring.
Crucially, for the purposes of transmission modeling, all patients in the
facility were tested for CRKP every other week after they were admitted. This
prospective surveillance program, along with the usual screening upon intake,
amounted to near-complete observation of cases (\cite{hawken2022threshold}).
Patients were relatively immobile during treatment, so we have detailed
knowledge of where patients were residing (i.e. facility floor and room) during
their stays, making it possible to model spatially heterogeneous transmission
risks through patient \textit{contact networks}.

Given its unusually dense sampling of cases, this dataset is atypical. Most HAIs
are asymptotic colonizers, so within-facility transmission often goes
undetected; prospective surveillance of pathogens is too costly to implement
universally. Modeling the epidemiological dynamics of these partially observed
systems at the individual level comes with computational and statistical
challenges, as we shall see in Section~\ref{sec:si-model}. For our data, fitting
a mechanistic transmission model via likelihood-based inference is relatively
straightforward, since individual-level latent variables are not needed. This
observational study is therefore a valuable benchmark for testing the efficiency
and accuracy of simulation-based calibration techniques such as ABC and NPE that
may be used on large-scale stochastic transmission models where standard
likelihood-based inference is inadequate.

\subsection{CRKP Dataset}
\label{sec:crkp-data}

In addition to regular CRKP screening and test results, the dataset includes the
time of admission and discharge for each patient as well as the floor(s) and
room(s) in which they stayed during their visit. We refer to time-varying
location information as the patient trace data. The dataset tracks 890 patients
and 1,112 distinct visits in total, with 18\% of patients making two or more
visits to the facility. Patients resided in five floors\footnote{We treat the
Special Care Unit, or SCU, as a floor, consistent with how the trace data are
encoded.} and 95 rooms; in many cases, patients move between multiple floors and
rooms over the course of one visit. The median length of a single visit was 24
days, while less than one in seven visits were shorter than a week. 95\% of
patients were tested within the first three days of their stay. We assume that a
positive test within three days of admission means a patient was colonized with
CRKP outside the facility and that CRKP could not be acquired from another
patient until at least the fourth day of a visit. 259 patients (29\%) tested
positive for CRKP at some point. Of these, 168 (65\%) imported CRKP into the
facility, whereas 91 (35\%) acquired CRKP after admission.

\begin{figure}
    \centering
    \begin{subfigure}[t]{0.45\textwidth}
        \centering
        \includegraphics[width=\textwidth]{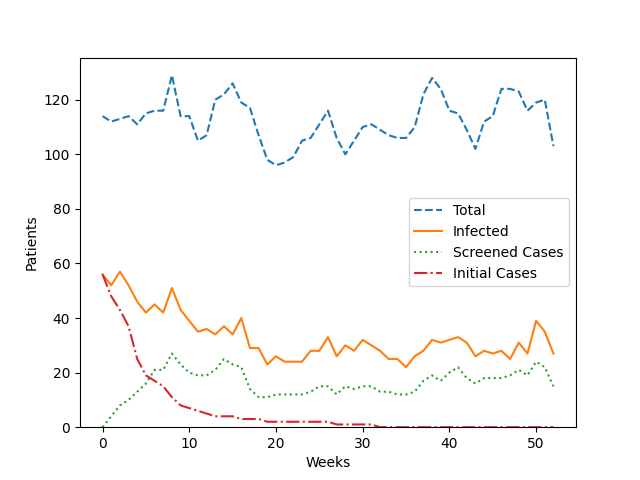}
        \caption{Breakdown of LTACH population over time.}\label{fig:crkp-viz}
    \end{subfigure}
    \begin{subfigure}[t]{0.45\textwidth}
        \centering
        \includegraphics[width=\textwidth]{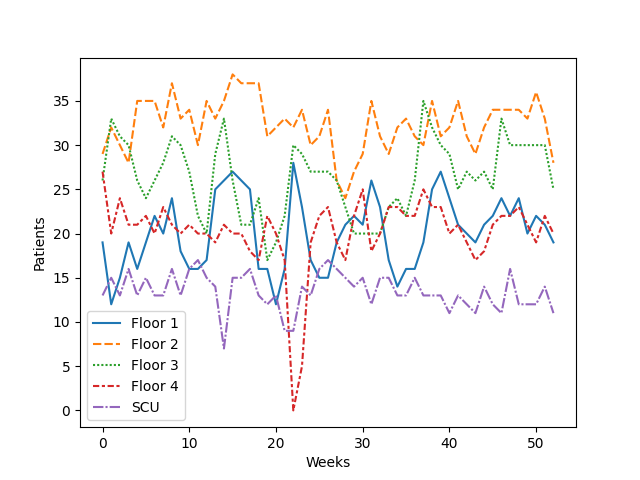}
        \caption{LTACH patient population by floor.}\label{fig:crkp-floors}
    \end{subfigure}
    \caption{Visualizing the LTACH CRKP Dataset.}
\end{figure}

Though the original data are recorded on a day-by-day frequency, we resample to a
weekly resolution. This smooths and reduces the dimensionality of the data
without erasing fluctuations over smaller timescales, since the vast majority of
patient stays last longer than a week. In Figure~\ref{fig:crkp-viz}, we
visualize a breakdown of the population of the LTACH over time. The overall patient
population varies between roughly 100 and 130 patients per week, 
with around 20 new patients admitted per week. About 50\% of patients are colonized with CRKP at
the start of the study, and that proportion decreases somewhat over the course
of the study, suggesting that the intervention was effective, consistent with
the assessment of the team carrying out the program
(\cite{hayden2015prevention}). We also highlight the cohort of patients who were
infected at the start of the study; by week 10, these patients have been mostly
discharged. Lastly, we show the number of patients who screen positive for CRKP
upon admission during the study. For most of the period of observation, the
majority of CRKP cases are importations from outside the facility, implying that
within-facility transmission contributes less to the overall incidence rate. We
show a breakdown of patient population by floor in Figure~\ref{fig:crkp-floors}.
Floor populations are generally stable, but the population of floor 4
temporarily plummets to zero around week 20. Given the coinciding spike in
population of Floor 1, it is possible that this is an issue of data sanity,
though the overall facility population appears to drop around week 20.

\section{Simulation-based Inference}\label{sec:sbi} 

Let $\bx_o$ be an observed dataset that depends on an unobserved set of 
parameters $\bth$ through a likelihood $p(\bx_o \mid \bth).$ Suppose that $\bth$ follows a prior
distribution $p(\bth).$ The primary aim of Bayesian inference is to compute the posterior distribution of
$\bth$ conditional on $\bx_o$. From Bayes' rule, it follows that
\begin{equation}
p(\bth \mid \bx_o) \propto p(\bth)p(\bx_o \mid \bth).
\end{equation}
Classical computational methods for Bayesian inference (e.g. MCMC) estimate the 
normalizing factor of the right-hand side of this expression through repeated evaluation 
of the joint density of the prior and likelihood. 

A \textit{simulator} is a computer program that
takes parameters $\bth$ as inputs, samples a sequence of random internal states
(latent variables) $\bz_t \sim p(\bz_t \mid \bth, \bz_{<t}),$ and finally
generates data $\bx \sim p(\bx \mid \bth, \bz)$ as output. If the simulation's latent space 
(representing the unobserved stochastic evolution of the system) is large, then the ensuing 
likelihood
\begin{equation}\label{eq:latent}
    p(\bx \mid \bth) = \int p(\bx, \bz \mid \bth) d\bz
\end{equation}
is a high-dimensional integral, and its direct evaluation is (perhaps prohibitively) expensive. However, 
sampling simulated data $\bx$ from the simulator is relatively easy.

This property of simulators motivates Approximate Bayesian Computation (ABC),
the most established method for simulation-based (a.k.a. ``likelihood-free'')
Bayesian inference (\cite{sisson2018handbook}). ABC compares many simulated
datasets to the observed data, rejecting parameters for simulations lying outside
an $\epsilon$-ball of the observed data for some error threshold $\epsilon > 0$.
ABC produces a nonparametric Monte Carlo sample that asymptotically approaches
the exact posterior. As $\epsilon$ decreases,
the ABC posterior approximation tends to improve, however, this increases the
procedure's computational cost, since more simulations are likely to be
rejected. ABC may scale poorly to high-dimensional models, as the number of
samples required to sufficiently explore the model space increases exponentially
(in the worse case) with the dimension of $\bth$ and $\bx.$ Low-dimensional summary statistics
of the data may improve sampling efficiency, but these can be hard to design and
will generally discard information. More sophisticated variants of ABC, such as
Sequential Monte Carlo ABC (SMC-ABC, see \citeauthor{toni2009approximate})
address certain limitations of standard ABC yet introduce further complexities, 
such as the need to tune
multiple $\epsilon$ thresholds.


Like ABC, NPE approximates the posterior distribution without directly
evaluating the likelihood density. Unlike ABC, NPE achieves this by fitting a
neural conditional density estimator $q_\phi(\cdot \mid \cdot)$ to the
density of $\bth$ conditional on $\bx$ for arbitrary $\bth$ and $\bx$ pairs sampled
from the joint model distribution $p(\bth, \bx)$. NPE performs \textit{amortized
inference}, exploiting information it learns about the ``general'' posterior to predict the 
specific posterior
conditional on $\bx_o$. For tractability, $q_\phi$ is constrained to belong to a
parametric family of densities; $\phi$ defines a neural network encoder function
mapping data to the conditional density of the model parameters. The encoder is
trained by maximizing the objective function
\begin{equation}\label{eq:npe-obj}
    \mathbf{E}_{p(\bth, \bx)}[\log q_\phi (\bth \mid \bx)]
\end{equation}
with respect to $\phi,$ using Stochastic Gradient Descent or a variant thereof.
We outline NPE in Algorithm~\ref{alg:npe}. 

\begin{algorithm}
    \caption{Amortized Neural Posterior Estimation}\label{alg:npe}
    \begin{algorithmic}
    \Require Prior distribution $p(\bth),$ simulator $p(\bx \mid \bth),$ neural
    conditional density estimator $q_{\phi}(\cdot \mid \cdot),$ observed data
    $\bx_o$ \Ensure Approximate posterior distribution $\hat p(\bth \mid \bx_o)$
    \For{$s = 1, 2, \ldots S$} \State Sample $\bth_s \sim p(\bth)$ \State
    Simulate $\bx_s \sim p(\bx \mid \bth_s)$ \EndFor \State Using stochastic
    gradient descent, solve
    \begin{equation}\label{eq:npe-emp}
    \phi^* = \argmin_{\phi} \quad - \frac{1}{S} \sum_{s=1}^S \log q_\phi(\bth_s \mid \bx_s)
    \end{equation}
    \State $\hat p(\bth \mid \bx_o) \gets q_{\phi^*}(\bth \mid \bx = \bx_o)$
    \end{algorithmic}
\end{algorithm}

NPE converts the problem of sampling from the posterior into an optimization
problem, analogous to standard, likelihood-based Variational Inference (VI), for which
the evidence lower bound (ELBOW) is the optimization objective. \citeauthor{ambrogioni2019forward} showed that maximizing
Equation~\ref{eq:npe-obj} is equivalent to minimizing the forward KL-divergence
between the model distribution $p(\bth, \bx)$ and the joint variational
distribution $q(\bth, \bx) = q_\phi(\bth \mid \bx)k(\bx),$ where $k(\bx)$ is the
sampling distribution of the simulated training data used to fit $q_\phi.$ 
NPE is part of a family of algorithms that use
neural networks trained on simulated data to approximate a posterior, known as
simulation-based inference (SBI, \cite{tejero-cantero2020sbi}). 

Once the encoder network has been trained, it is simple to produce an
approximate posterior conditional on any candidate dataset. The quality of this
approximation will strongly depend on how similar the target observed data are
to simulated datasets encountered during training. If the simulation model
$p(\bth, \bx)$ is an unfaithful descriptor of reality, then none of the
simulations $\bx$ will resemble $\bx_o$, no matter how large the simulation
budget. NPE and related neural SBI methods may yield unreliable inferences under
this scenario of \textit{model misspecification}, which corresponds to
out-of-distribution prediction, a notorious vulnerability of deep learning (\cite{ward2022robust}).
Posterior predictive checks, conservative priors, and judicious use of
dimension-reducing summary statistics can help mitigate the risk of model
misspecification, which we consider with respect to our application in
Section~\ref{sec:disc}.

In our experiments, we trained NPE to learn transformed Gaussian posterior approximations,
which are easy to interpret, optimize, and draw samples from. We deemed this choice of
conditional density estimator sufficiently flexible for our purposes since the models we considered
yielded unimodal and relatively narrow posteriors. For estimating more irregular posteriors,
Normalizing Flows are perhaps the most popular conditional density estimator in SBI (\cite{papamakarios2021normalizing}). 
We employed three-layer feedforward neural networks as our model architecture, with the
network width, training batch size, and weight decay regularization
(\cite{loshchilov2017fixing}) as the hyperparameters to be tuned. To prevent
overfitting, we used a 75-25 training/validation split of simulated data along
with early stopping.

\section{A Susceptible-Infected Model of Healthcare-Associated
Infection}\label{sec:si-model}

We develop a discrete-time, stochastic compartmental model with heterogeneous
infection rates for modeling healthcare-associated infections such as CRKP.
Through a simulation experiment, we compare the efficiency and accuracy of
likelihood-based and simulation-based techniques for model calibration.

A Susceptible-Infected (SI) model is appropriate for representing HAI transmission,
since colonized patients tend to remain colonized, barring medical intervention,
and patients do not acquire immunity after being colonized. Hospitals and other
care facilities are generally small-population settings featuring demographic
stochasticity (Figure~\ref{fig:crkp-floors}) which means that aggregate disease
spread appears non-deterministic (Figure~\ref{fig:crkp-viz}). Patients in our
target dataset are tested for infection at relatively frequent and regular
intervals, making a discrete-time model appropriate for describing transmission.

\subsection{SI Model with Heterogeneous Infection Rates}\label{sec:homo}

Suppose we observe a healthcare facility with a population of $N$ patients over
$T$ time steps. For each individual $i=1, \ldots, N,$ let the binary variable
$X^{(i)}_t$ represent their disease status at intervals $t = 1, \ldots, T.$ If a
patient is infected by time $t,$ then $X^{(i)}_t = 1,$ otherwise $X^{(i)}_t =
0.$ Let $S_t$ and $I_t$ denote the number of susceptible and infected patients
in the facility at time $t.$ Patients are either susceptible or infected, so
$S_t + I_t = N$ for all $t.$

We model the shift of patients from susceptible to infected stochastically. Let
$\lambda_i(t)$ denote the hazard acting on a susceptible individual $i$ at time
step $t$. We assume $\lambda_i$ is constant over each discrete interval $[t-1,
t)$ and that each individual infection event is an exponentially-distributed
event with rate parameter $\lambda_i(t)$. Over small enough time steps, this is
a reasonable assumption (\cite{king2017simulation}). Then,
\begin{equation}
    P(X^{(i)}_t \mid X^{(i)}_{t-1} = 0) = (1 - e^{- \lambda_i(t)})^{X^{(i)}_t}( e^{- \lambda_i(t)}) ^ {1 - X^{(i)}_t}.
\end{equation}
This formulation results in binomial sampling of cases. Each individual's
trajectory $\mathbf{X}^{(i)} = \{ X^{(i)}_1, \ldots, X^{(i)}_T \}$ is a
stochastic process: a discrete-time, non-homogeneous Markov Chain. We
can think of $\lambda_i(t)$ as the \textit{force of infection} acting on patient
$i$ at time $t.$


\subsubsection{Individualized Force of Infection}

We allow the force of infection to vary based on the spatial proximity of
possible donor (infectious) patients to a given recipient (susceptible) patient,
resulting in individualized transmission risks. Suppose that the facility
comprises $P$ floors and $R$ rooms. We consider a vector of heterogeneous
infection rates, $\bbeta = (\beta_0, \beta_1, \ldots, \beta_P, \beta_{P+1}),$
where each $\beta_j > 0$. These rates are interpreted as follows:
\begin{itemize}
    \item $\beta_0$ is the \textit{facility rate}, the number of
    transmission-producing contacts (secondary infections) an infectious patient
    makes with anyone in the facility per unit of time;
    \item For $p = 1, \ldots, P,$ $\beta_p$ is the \textit{floor rate}, the
    number of transmission-producing contacts an infectious patient in floor $p$
    makes with a floormate per time unit;
    \item $\beta_{P+1}$ is the \textit{room rate}, the number of
    transmission-producing contacts an infectious patient makes with a roommate
    per time unit.
\end{itemize}
Let $F(i) \in \{1, \ldots, P\}$ denote the floor where patient $i$ is staying at
time step $t$. Let $N_F(i)$ denote the population on floor $F(i)$, and let $N_R$
be the maximum number of patients in a room. The individual contribution of an
infected patient $j$ to the force of infection acting on an susceptible patient $i$
is
\begin{equation}
    \lambda_{i \leftarrow j} = \begin{cases}
        \dfrac{\beta_0}{N} + \dfrac{\beta_{F(i)}}{N_{F(i)}} + \dfrac{\beta_{P+1}} {N_R} &{\text{if $i$ and $j$ are roommates,}} \\
        \dfrac{\beta_0}{N}+ \dfrac{\beta_{F(i)}}{N_{F(i)}} &{\text{if $i$ and $j$ are floormates,}} \\
        \dfrac{\beta_0}{N}&{\text{otherwise.}}
    \end{cases}
\end{equation}
For each spatial level, we divide the location-specific transmission rate by the
local subpopulation size, which yields the risk of contagious contact per
individual recipient. The aggregate force of infection acting on an individual
is then
\begin{equation}\label{eq:het-hazard}
    \lambda_i(t) = \sum_{j: X_{t-1}^{(j)} = 1} \lambda_{i \leftarrow j}.
\end{equation}
The individual force of infection is a linear expression depending on the
recipient patient's contact network at the facility, floor, and room level.
All $\beta_j$ are positive, so this formulation assumes that the pairwise infection
risk increases additively with the proximity of a donor and recipient. One
justification for this modeling choice is the plausible tendency of healthcare
workers, the likeliest transmission vectors, to preferentially associate with
patients in the same location.

Homogeneous transmission, a.k.a. random mixing, is a special case of this
heterogeneous model. If we assume that the location of a potential infected
donor with respect to a susceptible recipient does not matter, then the force of
infection depends only on the facility-level infection rate, $\beta = \beta_0$.
Therefore,
\begin{equation}
    \lambda_i(t) = \sum_{j: X_{t-1}^{(j)} = 1} \beta = \beta \frac{I_{t-1}}{N}.
\end{equation}

\subsubsection{Intake and Outtake}
\label{sec:random-turnover}

Healthcare facilities see rapid turnover of patients relative to the timescale
of disease transmission. In real-world studies of HAIs, we can expect to have
information on patient admission and discharge times and possibly whether they
imported infection into the facility. We treat admission and discharge as fixed
events when modeling transmission in the CRKP dataset
(Section~\ref{sec:application}), but for the purpose of simulation experiments
we adopt a random patient turnover model for intake and outtake.

Suppose that at any time step, patients are discharged with a fixed probability
$\gamma$ and immediately replaced, with the facility remaining at full capacity
at all times. Let $\alpha$ denote the population proportion of already-colonized
individuals entering the facility. We can think of the sample index $i = 1,
\ldots, N$ as referring to individual \textit{locations} (e.g. patient beds)
that host patients as they move in and out of the facility. We assume that
$\alpha$ and $\gamma$ are known quantities, unlike the infection
rate(s).\footnote{The resulting model is a stochastic counterpart to the
deterministic Ross-Macdonald model (\cite{ross1911prevention,
macdonald1957epidemiology}), a classical model for the transmission of
mosquito-borne malaria. Here, mobile healthcare workers act as the disease
vector between stationary patients (\cite{doan2014optimizing}).}

The status $X_t^{(i)}$ of the patient at location $i$ and time $t$ may change
due to one of three random events. We list these transition events along with
their probabilities:
\begin{enumerate}
    \item $P(\text{a patient is replaced by an infected}) = \gamma \alpha$
    \item $P(\text{a patient is replaced by a susceptible}) = \gamma (1 -
    \alpha)$
    \item $P(\text{a \textbf{susceptible} patient $i$ is infected}) = (1 -
    \gamma) \cdot (1 - e^{- \lambda_i(t)})$
\end{enumerate}
We assume that a susceptible patient cannot get infected until the time step
after they are admitted. Therefore, a freshly admitted patient is immune from
infection, explaining the $(1 - \gamma)$ term in event 3. We write out the data
generating process as a simulation algorithm in Supplement~\ref{supp:si-algorithm}.

\subsubsection{Model Likelihood}

Given an observed dataset $\bX$ of patient statuses over time, we can write out
the likelihood of a vector of infection rates $\bbeta$. Let $X_t = (X^{(1)}_t,
\ldots, X^{(N)}_t)$ denote the array of patient statuses at time $t.$ At the
start of the observation period ($t=1$), we sample initial statuses from a
Bernoulli with probability $\alpha,$ the population proportion of infection.
Then,
\begin{equation}\label{eq:likelihood-base}
    P(X_1) = \prod_{i=1}^N \alpha^{X^{(i)}_1} (1 - \alpha)^{1 - X^{(i)}_1}
\end{equation}
For subsequent time steps $t=2, \ldots T,$ we can write an autoregressive
conditional probability as
\begin{equation}\label{eq:likelihood-step}
\begin{split}
    P(X_t \mid X_{t-1}) = \prod_{i=1}^N \left ( [\gamma \alpha + (1 - \gamma)]^{X^{(i)}_{t-1}} \cdot [\gamma \alpha + (1 - \gamma) (1 - e^{-\lambda_i(t) })]^{(1 - X^{(i)}_{t-1})} \right )^{X^{(i)}_{t}} \\
    \cdot \left ( [\gamma (1 - \alpha) ]^{X^{(i)}_{t-1}} \cdot [\gamma(1 - \alpha) + (1 - \gamma) (e^{- \lambda_i(t) }) ]^{(1 - X^{(i)}_{t - 1})} \right ) ^ {(1 - X^{(i)}_{t} )}.
\end{split}
\end{equation}
Let $\bX = X_1, \ldots, X_T.$ The complete likelihood over all time steps then
factorizes as 
\begin{equation}\label{eq:likelihood-full}
    p( \bX \mid \beta)  =  P(X_1) \cdot \prod_{t=2}^T P(X_t \mid X_{t-1}).
\end{equation}
See Supplement~\ref{supp:si-derivation} for a derivation of the likelihood.

Evaluating the likelihood has a fairly tractable computational complexity of
$O(N^2)$. Suppose, however, that at any time step, we observe only a fraction of
cases, a common problem in the study of HAIs, which are often asymptomatic
colonizers up until the point of invasive infection. Under partial observation
of cases and heterogeneous transmission effects, modeling unknown patient
statuses necessitates using individual-level latent variables. Marginalizing
over these latents to recover the complete likelihood has an exponential
complexity of $O(2^N \cdot N^2)$, so conventional likelihood-based inference
methods will struggle to scale to all but the smallest study population sizes.
We explore this scenario in Supplement~\ref{supp:sim-figures}.

\subsection{Simulation Experiment: Heterogeneous Transmission}
\label{sec:hetero}

In this experiment, we assess the efficiency and accuracy of simulation-based
calibration for a HAI transmission model where the infection rates
$\bbeta$ vary across five floors and between roommates. For our ``observed''
data, we simulated an outbreak using known parameter values (see
Table~\ref{tab:het-means}); visualization of the
location-specific incidence can be found in Supplement~\ref{supp:sim-figures}.  We set the remaining simulation
parameters to be $N=300$, $T=52$, $\alpha = 0.1$, and $\gamma = 0.05.$
We assign $\bbeta$ a multivariate lognormal prior
with a diagonal (independent components) covariance structure:
\begin{equation}
    \log(\bbeta) \sim \mathcal{N}\left (\boldsymbol{\mu} = [-3, -4, -4, -4, -4, -4, -4], \boldsymbol{\Sigma} = I_7 \right ).
\end{equation}
By assigning a lower prior mean to the sublocation rates (within-floor and
within-room infection), this choice of prior reflects a cautious belief that
defaults back on a mainly homogeneous model of transmission in the absence of
evidence to the contrary. As a benchmark for calibration accuracy, we used emcee (\cite{foreman2013emcee}),
an implementation of the Affine Invariant MCMC sampler, to obtain a likelihood-based estimate of the posterior distribution $p(\bbeta \mid \bX)$. To
achieve sufficient mixing, we ran 16 chains with 2,000 steps each, for a total
of 32,000 evaluations of the likelihood.

\begin{table}
\caption{Posterior mean point estimates of heterogeneous infection rates.}
\centering
\begin{tabular}{lllll}
\toprule
Transmission Rate & Value & MCMC & NPE & ABC \\
\midrule
Facility & 0.05 & 0.0645 & 0.0800 & 0.0882 \\
Floor 1 & 0.02 & 0.00994 & 0.0119 & 0.0146 \\
Floor 2 & 0.04 & 0.0465 & 0.0282 & 0.0270 \\
Floor 3 & 0.06 & 0.0240 & 0.0221 & 0.0199 \\
Floor 4 & 0.08 & 0.0458 & 0.0367 & 0.0301 \\
Floor 5 & 0.1 & 0.0926 & 0.0792 & 0.0496 \\
Room & 0.05 & 0.0458 & 0.0278 & 0.0291 \\
\bottomrule
\end{tabular}
\label{tab:het-means}
\end{table}

Table~\ref{tab:het-means} shows the estimates of the heterogeneous infection
rates given by MCMC, NPE, and ABC. We allocated a simulation budget of 16,000
samples to ABC and NPE. We summarized the data by combining seven descriptive
statistics or ``views'' of the data, each a time series corresponding to one of
the location-specific parameters $\beta_j$ in $\bbeta$.\footnote{Standard neural
network architectures are not permutation invariant, so pooling raw data across
observation indices, e.g. by summation, is commonly used to preprocess NPE
inputs. An alternative approach is to use special set-based architectures
 for the encoder (\cite{chan2018likelihood}).} These statistics are: the total
case rate over time (the proportion infected, $\bI$), the case rates by floor,
and the time-varying proportion of rooms with both roommates infected. We denote
these statistics collectively as $\mathbf{J},$ which is a high-dimensional
vector of length 364 (7 by 52). In Supplement~\ref{supp:summ-stats}, we provide details on how
$\bJ$ is computed from $\bX$ and the floor and room traces.

The NPE estimates of all infection rates are more accurate than the ABC
estimates, with the exception of the room-level infection rate, which is a
virtual tie between the two. The ABC estimates of the sublocation rates appear
to be biased towards the prior (proposal) mean. ABC is in theory an
asymptotically unbiased estimator of the posterior, but this is not a practical
guarantee for finite simulation budgets. In contrast, NPE seems
better able to pick up on signal from the data for the floor-level rates given
the same number of simulations. However, room-level transmission, as captured
through our summary statistic, may be too noisy a phenomenon for either NPE or
ABC to estimate very precisely. We provide additional metrics of the relative 
performance of ABC and NPE in Supplement~\ref{supp:sim-figures}.

We plot the efficiency-accuracy trade-off
for three simulation-based posterior estimators in Figure~\ref{fig:het-error}, using our likelihood-based estimate 
as an accuracy benchmark. We compare the
performance of ABC, NPE with a full-covariance multivariate normal (MVN)
approximation of $\log(\bbeta)$, and NPE with a diagonal covariance MVN, also on
the log scale. The latter assumes a fully factorized posterior with independent
components, the so-called mean-field approximation, so we denote it as NPE-MF.
By ignoring posterior correlations, NPE-MF has fewer parameters to estimate as
compared to full-covariance NPE (14 vs. 35 for a seven-dimensional variable) at
the cost of producing a less informative and perhaps more biased estimate.

\begin{figure}
\begin{subfigure}[t]{.45\textwidth}
    \centering
    \includegraphics[width=\linewidth]{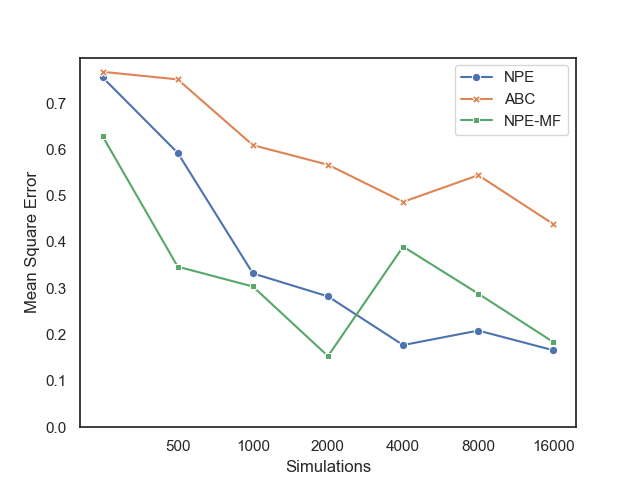}
    \caption{Aggregated squared error across all seven $\beta_j$ in $\bbeta$.}
    \label{fig:het-error-a}
  \end{subfigure}
  \hfill
  \begin{subfigure}[t]{.45\textwidth}
    \centering
    \includegraphics[width=\linewidth]{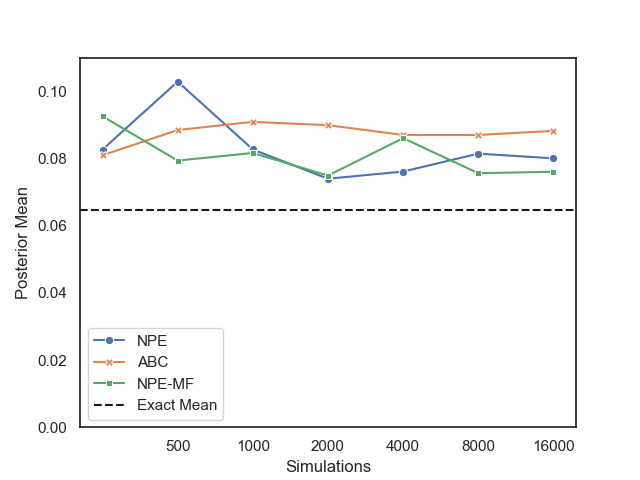}
    \caption{Estimation of the facility infection rate $\beta_0$.}
    \label{fig:het-error-b}
  \end{subfigure}

  \medskip

  \begin{subfigure}[t]{.45\textwidth}
    \centering
    \includegraphics[width=\linewidth]{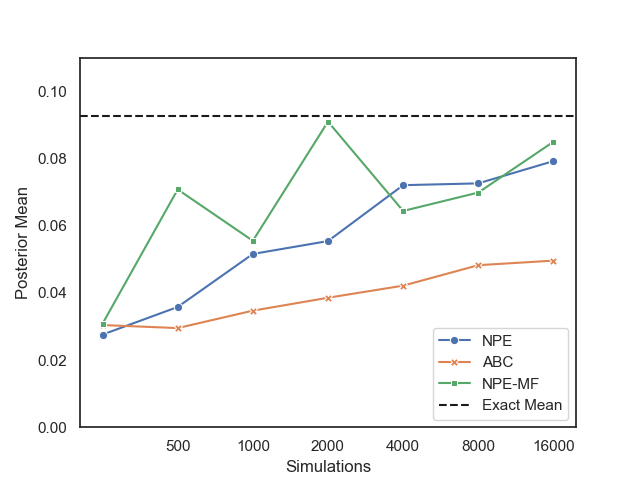}
    \caption{Estimation of the infection rate within Floor 5, $\beta_5$.}
    \label{fig:het-error-c}
  \end{subfigure}
  \hfill
  \begin{subfigure}[t]{.45\textwidth}
    \centering
    \includegraphics[width=\linewidth]{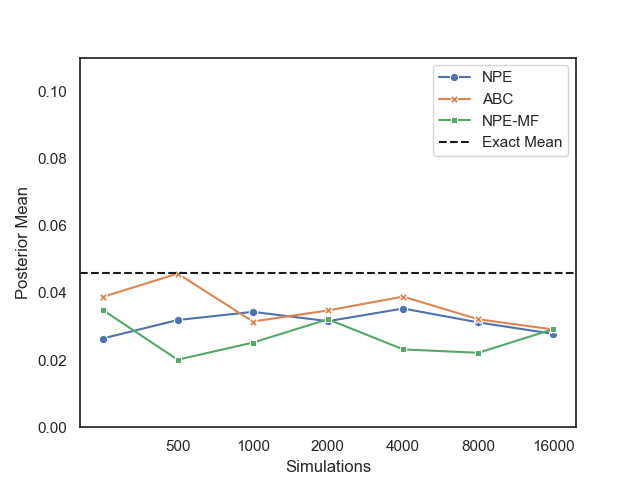}
    \caption{Estimation of the infection rate between roommates, $\beta_6$.}
    \label{fig:het-error-d}
  \end{subfigure}
\caption{Simulation-based estimation accuracy and sample-efficiency for heterogeneous infection rates. Likelihood-based posterior mean estimates are used as the baseline.}
\end{figure}

We varied the size of the simulation budget from 250 to more than 16,000
samples. In Figure~\ref{fig:het-error-a}, we show the mean squared error across
all 7 parameters on the log scale. As a representative subset of the 7
components of $\bbeta$, we also plot the convergence of the estimates for the
the facility-level  transmission rate $\beta_0$ (Figure~\ref{fig:het-error-b}),
the rate within floor 5 $\beta_5$ (Figure~\ref{fig:het-error-c}), and the
room-level rate $\beta_6$ (Figure~\ref{fig:het-error-d}). (For the remaining
components, see Supplement~\ref{supp:sim-figures}.) For smaller simulation budgets, NPE-MF converges
more rapidly on the exact posterior mean than NPE, generally requiring half as
much training data to produce a point estimate of $\bbeta$ with similar
accuracy. This is expected behavior, as NPE-MF has fewer parameters to estimate.
In contrast, NPE does not appear to converge on a solution until 4,000 training
samples, at which point the error of NPE-MF spikes upward. Eventually, the
NPE-MF aggregate error decreases to the same level as full-covariance NPE.

Compared to NPE and NPE-MF, ABC displays worse efficiency and accuracy,
requiring orders of magnitude more simulations to achieve an inferior estimate
of the parameters. In particular, the ABC estimate of $\beta_5$, the true value
of which is relatively far from the prior, shows exhibits stubborn downward
bias. The performance gap between ABC and NPE is most pronounced for components
of $\bbeta$ with low prior density assigned to the true values. This is a
significant challenge for ABC, which relies on sampling enough nearly correct
points from the prior (a.k.a. proposal) distribution. In contrast, NPE is a
regression technique that can interpolate the observed posterior from relatively
scarce simulated data.

\label{fig:het-error}

\begin{figure}
  \centering
  \includegraphics[width=0.45\linewidth]{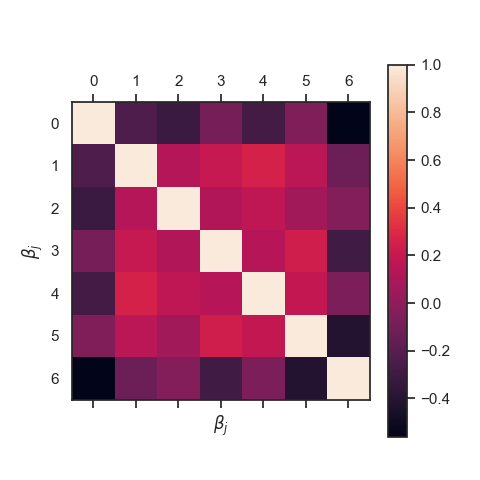}
  \caption{Correlation heatmap of heterogeneous infection rates $\beta_j$ ($j = 1, \ldots 6)$, estimated via NPE using a MVN approximation.}
  \label{fig:corr}
\end{figure}

In Figure~\ref{fig:corr}, we show a correlation heatmap for the NPE posterior
estimate. The five floor-level rates are positively
correlated with one another and negatively correlated with the facility and
room-level rates. The facility and room-level rates show a particularly strong
negative correlation with one another. We interpret negative posterior
correlation between model parameters as indicating competing explanations of the
observed data: for example, high facility-wide incidence levels could be
explained by a large facility infection rate or by a high rate of transmission
between roommates across the facility. NPE identified a covariance structure
similar to the one found by MCMC (see Supplement~\ref{supp:sim-figures}). We attribute the erratic convergence behavior
of NPE-MF to its inability to model significant posterior correlations between
infection rates, though it is possible we did not explore a sufficiently wide
range of hyperparameter settings.

\section{Application to CRKP LTACH Transmission}\label{sec:application}

\subsection{CRKP Transmission Model}
\label{sec:crkp-model}

Turning now to analysis of CRKP transmission in a LTACH, we adapt the stochastic
SI model introduced in the previous section. As before, we assume that patients
are either uncolonized with CRKP or colonized (infected) and that no recovery
occurs. We also assume that there are no unobserved cases. In our simulation
experiments, we modeled turnover of patients and importation of new cases
randomly (Section~\ref{sec:random-turnover}), but our dataset gives us knowledge
of the exact times that patients enter and leave the LTACH as well as their
screening test results for CRKP upon arrival. Therefore, we treat patients'
visit times and imported infections as fixed events. Our model simulates
within-facility acquisition of CRKP based on a time-varying and possibly
spatially-dependent force of infection. 

The CRKP study tracks $N = 890$ distinct patients over the course of $T = 53$
weeks. As before, let $\bX$ be an $N \times T$ matrix of patient infection
status, with $X_t^{(i)}$ equaling one if patient $i$ is infected with CRKP at
the start of week $t$ and zero otherwise. We let the $N \times T$ matrices
$\mathbf{W}$, $\mathbf{F}$, and $\mathbf{R}$ denote the facility trace, floor
trace, and room trace respectively. For the facility trace, $W_t^{(i)}$ equals
one if patient $i$ is present in the facility during week $t$ and zero
otherwise. Lastly, let  $\mathbf{V}$ be the record of screening results: for a
patient $i$ newly admitted to the facility at time $t$, $V_t^{(i)}$ equals one
if the patient tests positive for CRKP during screening and zero if that patient
has a negative result. We write out the simulation process for generating data
in Algorithm~\ref{alg:crkp}. Since this process is fully observed, evaluation of
the model's likelihood is computationally tractable: we provide a derivation of
the likelihood in Supplement~\ref{supp:crkp-model}.

\begin{algorithm}
    \caption{CRKP Transmission Simulator}\label{alg:crkp}
    \begin{algorithmic}
        \Require Vector of transmission rates $\boldsymbol{\beta}$, facility
        trace $\mathbf{W}$, floor trace $\mathbf{F}$, room trace $\mathbf{R}$,
        screening results $\mathbf{V}$ \Ensure $N \times T$ matrix $\mathbf{X}$
        of infection statuses $\mathbf{X}^{(i)} = \{X^{(i)}_1 \ldots X^{(i)}_T
        \}$ for all patients $i=1, \ldots, N$ \For{$t = 1, 2, \ldots T$} \For{$i
        \in 1, \ldots, N$} \If{$W_t^{(i)} = 1$ and $W_{t-1}^{(i)} = 0$ (or $t =
        1)$} \Comment{$i$ is newly admitted} \State Set $X_t^{(i)} = V_t^{(i)}$
        \Comment screening results are fixed \Else \If{$W_t^{(i)} = 1$ and
        $W_{t-1}^{(i)} = 1$} \Comment{$i$ has stayed at least one week}
        \If{$X_{t-1}^{(i)} = 0$} \Comment $i$ is susceptible \State Compute the
        individualized force of infection $\lambda_i(t)$
        (Eq.~\ref{eq:het-hazard})\State Draw $X_t^{(i)} \sim \mbox{Bernoulli}(1
        - e^{-\lambda_i(t)})$ \Else \State $X_t^{(i)} \gets 1$ \EndIf \Else
        \State Set $X_t^{(i)} = \varnothing$ \Comment $i$ is not present at time
        $t$ \EndIf \EndIf \EndFor \EndFor
    \end{algorithmic}
\end{algorithm}

\subsection{Homogeneous Model Results}

\begin{figure}
    \centering
    \begin{subfigure}[t]{0.45\textwidth}
        \centering
        \includegraphics[width=\textwidth]{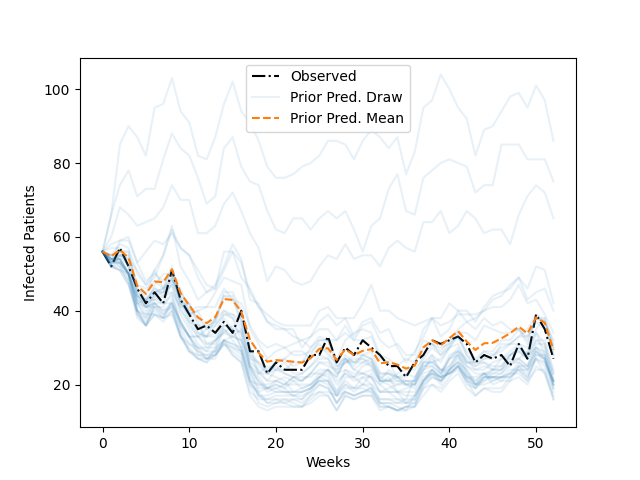}
        \caption{Prior predictive check of $\beta$.}\label{fig:crkp-prior-pc}
    \end{subfigure}
    \begin{subfigure}[t]{0.45\textwidth}
        \centering
        \includegraphics[width=\textwidth]{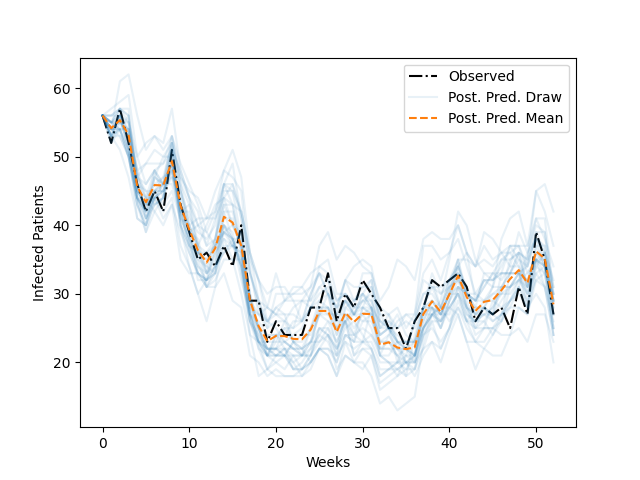}
        \caption{NPE Posterior predictive check of
        $\beta$.}\label{fig:crkp-ppc-homog}
    \end{subfigure}
    \caption{Predictive checks for the homogeneous CRKP transmission model}
\end{figure}

First, we fit a homogeneous model of CRKP transmission to the data. We assign
the prior $\beta \sim \text{Lognormal}(\mu = -2, \sigma = 1)$ to the
facility-wide rate of infection. A prior predictive check of total CRKP
incidence (Figure~\ref{fig:crkp-prior-pc}) suggests that this prior is
reasonable, though a few predictive draws greatly overshoot the observed data,
illustrating the unstable, nonlinear dynamics at play in our stochastic model. 

We trained NPE on 4,000 simulations of $\bI$ (incidence over time, see
Section~\ref{sec:homo}) to learn a lognormal approximation to the posterior of
$\beta$, $\hat p(\beta \mid \bI_o) = \text{Lognormal}(\mu = -2.06,
\sigma=0.120).$ This gives a point estimate for the homogeneous infection rate
of 0.129 secondary infections per week. This is slightly higher than 0.124
secondary infections per week, the estimated rate produced by both MCMC (8,000
evaluations of the likelihood) and ABC (7,858 simulation samples). The MCMC
posterior had slightly lower variance than both ABC and NPE, which we would
expect because MCMC estimated the posterior conditional on the raw data $\bX_o$
through the model likelihood, whereas ABC and NPE approximated the posterior
conditional on overall incidence, a non-sufficient summary statistic.

Posterior predictive checks are useful for validating inference when,
as in our application, the true parameter values are unknown. As a quantitative
metric of model calibration accuracy, we consider the mean squared posterior predictive
error, MSPPE. Let $g(\bX)$ be an arbitrary
summary statistic of the data. We then define MSPPE as the expectation
$\mathbf{E}_{p(g(\bX) \mid \bX_o)}\left [\norm{g(\bX) - g(\bX_o)}^2 \right ]$,
where $p(g(\bX) \mid \bX_o)$, the posterior predictive density, factorizes as $
p(g(\bX) \mid \bth)\cdot p(\bth \mid \bX_o).$ This expectation can be estimated
by averaging the error of simulation draws over the (estimated)
posterior.\footnote{MSPPE is an imperfect measure of model calibration accuracy.
First, the exact posterior will only be unbiased in large sample settings.
Second, underestimating the true posterior uncertainty may reduce MSPPE. Third,
MSPPE is an in-sample calibration metric and is not useful for detecting
overfit. MSPPE should be treated as a heuristic for comparing posterior
estimates.} MCMC, NPE, and ABC yielded similar MSPPE with respect to facility
wide incidence $\bI$: 19.8, 20.0, and 21.3 respectively, implying that these
three posterior estimates are comparably accurate.

We show a posterior predictive check of incidence for NPE in
Figure~\ref{fig:crkp-ppc-homog}; the calibrated model exhibits relatively low
aleatoric uncertainty, with most predictive draws falling a short distance from
the observed data. For all three estimated posteriors, the calibrated
homogeneous model is an unbiased predictor of facility-wide CRKP incidence, but
it overestimates the incidence on floors 2 and 4 while slightly underestimating
the incidence on floor 3. (We show the complete set of location-based posterior
predictive checks for the NPE homogeneous estimate in Supplement~\ref{supp:crkp-checks}.) While the
homogeneous model appears to be a realistic model of overall transmission of
CRKP, it is insufficiently expressive to capture local infectious dynamics. 

\subsection{Heterogeneous Model Results}

Next, we fit a heterogeneous transmission model to the data, allowing the
infection rate to vary within each of the five floors and between roommates. We
assign the following multivariate lognormal prior:
\begin{equation}
    \log(\bbeta) \sim \mathcal{N}\left ( \boldsymbol{\mu} = [-3, -4, -4, -4, -4, -4, -5], \boldsymbol{\Sigma} = I_7 \right ).
\end{equation}
This is a somewhat conservative prior that leans towards a mainly-homogeneous
transmission mechanism. As before, we compare a likelihood-based MCMC estimate
of the posterior to simulation-based NPE and ABC estimates. Both NPE and ABC
return an approximate posterior $p(\bbeta \mid \bJ)$, where $\bJ$ is made up of
seven summary time series, each corresponding to a location for a
susceptible-infected transmission pair (see Section~\ref{sec:hetero}). For MCMC,
we ran 16 chains with 5,000 steps each for a total of 80,000 evaluations of the
model likelihood. NPE trained on 5,000 simulations to learn a multivariate,
full-covariance lognormal approximation to the posterior, and ABC accepted 100
samples out of 12,503 simulation draws.

\subsubsection{Pairwise Relative Risks}

Table~\ref{tab:crkp-risk} shows the estimated pairwise relative risks of
infection by location for each posterior estimate (MCMC, NPE, and ABC).
Infection risks are effectively per capita infection rates (cf.
Equation~\ref{eq:het-hazard}).\footnote{For the unadjusted posterior rate
estimates see Supplement~\ref{supp:crkp-figures}.} We show risks as ratios with respect to the
facility-wide risk. ABC appears to estimate relative risks similar to those of
our prior, suggesting that a patient is slightly more likely to be infected by a
floormate than by an arbitrary infected patient in the facility and is 7-9 times
more likely to be infected by a roommate. The MCMC posterior estimate infers a
\textit{reduced} risk of infection between floormates than at the facility level
(except for floormates in the SCU), which implies primarily random mixing---that
is, homogeneous transmission---in the LTACH. NPE, however, predicts strongly
elevated infection risks within Floor 3 (6 times baseline facility-level risk)
and between roommates (22 times facility-level risk). In other words, NPE finds
that transmission is associated more strongly with the spatial proximity of
patients than the other posterior estimators.

\subsubsection{Posterior Variance}

As well as point estimates of infection risk, we also compare the three methods
in terms of their estimated posterior uncertainty, i.e. variance. All else being
equal, a strong signal in the data will correspond to decreased variance in the
posterior estimate. Table~\ref{tab:crkp-sd} shows the marginal standard
deviations for each of the infection rates as inferred by MCMC, NPE, and ABC.
The former is the only method to exploit full information via the unsummarized
likelihood, and so it shows lower posterior variance across the board. However,
MCMC appears to assign relatively high uncertainty to the floor- and room-level
infection rates. This, along with the computed relative risks, could be
interpreted as weak evidence of spatially heterogeneous transmission. NPE and
ABC make use of summarized simulated data $\bJ$, so it is unsurprising that they
yield higher posterior uncertainty than MCMC. NPE arrives at a tighter estimate
of the facility-wide infection rate $\beta_0$ than ABC, but for the remaining
sublocation rates both ABC and NPE are no more certain than the prior. It is
likely that the location-specific summary statistics are not very informative of
proximate infection risk.

\subsubsection{Posterior Predictive Checks}

\begin{figure}
    \centering
    \begin{subfigure}[t]{0.45\textwidth}
        \centering
        \includegraphics[width=\textwidth]{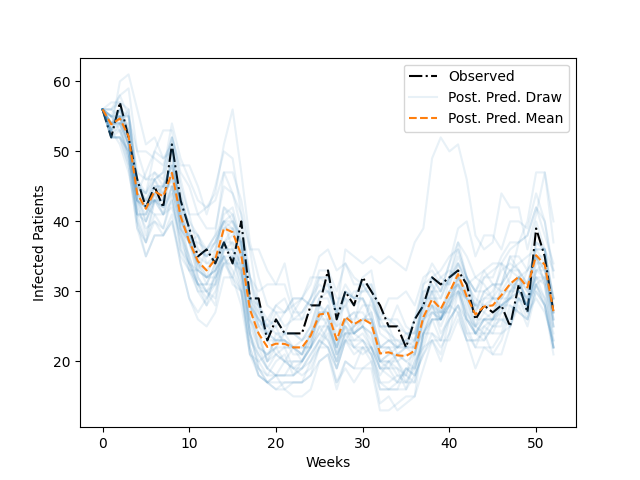}
        \caption{NPE posterior predictive check for facility-wide incidence.}\label{fig:crkp-ppc-npe}
    \end{subfigure}\hfill
    \begin{subfigure}[t]{0.45\textwidth}
        \centering
        \includegraphics[width=\textwidth]{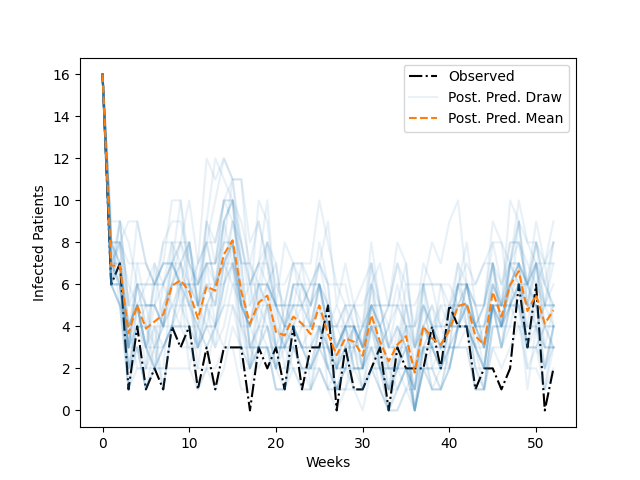}
        \caption{MCMC posterior predictive check for Floor 2 incidence.}\label{fig:crkp-ppc-mcmc}
    \end{subfigure}
    \caption{Selected posterior predictive checks for the heterogeneous CRKP transmission model.}
\label{fig:crkp-ppc-het}
\end{figure}

We provide the complete set of posterior predictive checks for all three estimators and all
seven location-based descriptive statistics in Supplement~\ref{supp:crkp-checks}; we highlight two
predictive checks in Figure~\ref{fig:crkp-ppc-het} that illustrate limitations
of both the NPE and MCMC estimates of $\bbeta$. Figure~\ref{fig:crkp-ppc-npe}
shows how the NPE-calibrated heterogeneous transmission model slightly
underpredicts the facility-wide incidence, particularly from weeks 20-40 of the
study. Figure~\ref{fig:crkp-ppc-mcmc} depicts how MCMC overestimates the
incidence on Floor 2, predicting a small spike around week 15 that never
occurred. It is possible that NPE estimates too low a value for $\beta_0$, but
it also appears that the mainly-homogeneous model inferred by MCMC is
unrealistic at the floor level.

We compare the MSPPE for each of the three estimators, along with the prior
predictive error as a baseline, in Table~\ref{tab:mspe}. NPE shows improved
error for the facility incidence over ABC despite its biased predictions
(Figure~\ref{fig:crkp-ppc-npe}). This can be explained by the fact that ABC's
posterior predictions are rather volatile and high variance. NPE achieves a more
accurate prediction of the incidence on Floors 2 and 4 compared to the other
inference methods but a slightly worse prediction of the Floor 3 incidence.
Though NPE performs its calibration using incomplete, summarized information,
its aggregate MSPPE is effectively tied with that of MCMC, demonstrating the
empirical predictive accuracy of deep learning. The MCMC and NPE estimates of
$\bbeta$ diverge substantially (see Table~\ref{tab:crkp-risk}), so the
heterogeneous transmission model may be weakly identifiable, such that multiple
estimates of $\bbeta$ explain the observed data equally well.

\subsubsection{Model Criticism}

None of the three calibrated models appear to fully capture the observed
transmission and incidence of CRKP by floor, judging by the posterior predictive
checks, so we surmise that our heterogeneous transmission model is
misspecified. Examining the movement of admitted patients between floors, we
found that a disproportionate number (68\%) of patients moved to Floor 3 were
already infected, and that many of these patients came from Floors 2 and 4. We
refer the reader to Supplement~\ref{supp:crkp-figures} for a summary of patient mobility between
floors. A number of infected patients were placed in isolation as part of the
LTACH's intervention program (\cite{hayden2015prevention}), and we speculate
that this isolation took place on Floor 3, which had the highest observed CRKP
incidence of any floor. Posterior predictive checks indicate that the NPE and
MCMC posteriors slightly underestimate the incidence on Floor 3, which may be the
result of relocation of infected patients rather than within-floor transmission.
Additionally, the MCMC estimate predicts small spikes in infection on Floors 2
and 4 that never factually occurred, which could be due to the systematic depletion
of the infected population on these floors. Our model does not
mechanistically account for non-random segregation of infected patients, which
would complicate the estimation of location-specific infection rates.

\begin{table}
\caption{Pairwise relative risk (RR) of infection, relative to facility-level}
\centering
\begin{tabular}{lrrrr}
\toprule
Transmission Zone & Prior & MCMC & NPE & ABC \\
\midrule
Facility & 1 & 1 & 1 & 1 \\
Floor 1 & 1.69 & 0.593 & 3.27 & 1.48 \\
Floor 2 & 1.25 & 0.994 & 1.81 & 0.924 \\
Floor 3 & 1.356 & 0.404 & 5.95 & 1.45 \\
Floor 4 & 1.76 & 0.625 & 2.53 & 1.20 \\
SCU & 2.79 & 2.32 & 4.46 & 2.27 \\
Room & 8.73 & 3.49 & 22.33 & 7.04 \\
\bottomrule
\end{tabular}
\label{tab:crkp-risk}
\end{table}

\begin{table}
\caption{Estimated marginal standard deviations (log-scale) of the heterogeneous CRKP infection rates}
\centering
\begin{tabular}{lrrrr}
\toprule
Rate & Prior & MCMC & NPE & ABC \\
\midrule
Facility & 1.0 & 0.115 & 0.345 & 0.450 \\
Floor 1 & 1.0 & 0.761 & 1.06 & 0.926 \\
Floor 2 & 1.0 & 0.964 & 0.934 & 0.981 \\
Floor 3 & 1.0 & 0.697 & 0.918 & 1.09 \\
Floor 4 & 1.0 & 0.800 & 0.930 & 0.958 \\
SCU & 1.0 & 0.828 & 0.961 & 1.01 \\
Room & 1.0 & 0.794 & 1.09 & 1.01 \\
\bottomrule
\end{tabular}
\label{tab:crkp-sd}
\end{table}

\begin{table}
\caption{Mean Squared Predictive Error}
\centering\begin{tabular}{lrrrr}
\toprule
Descriptive & Prior & MCMC & NPE & ABC \\
\midrule
Facility Incidence & 115 & 20.9 & 25.2 & 32.7 \\
Floor 1 Incidence & 7.65 & 4.14 & 4.57 & 4.29 \\
Floor 2 Incidence & 15.0 & 8.84 & 4.56 & 10.2 \\
Floor 3 Incidence & 13.0 & 7.70 & 8.36 & 6.36 \\
Floor 4 Incidence & 7.12 & 4.59 & 2.82 & 6.14 \\
Floor 5 Incidence & 3.66 & 2.38 & 2.53 & 2.32 \\
Infected Rooms & 24.2 & 11.4 & 11.7 & 11.0 \\
\bottomrule
\end{tabular}
\label{tab:mspe}
\end{table}

\section{Discussion}\label{sec:disc}

In our simulation experiment, NPE accurately approximates the likelihood-based
posterior for unknown model parameters and shows improved efficiency compared to
ABC and MCMC. NPE leverages key strengths of deep learning, including accurate
interpolation between points in high dimensional space and automatically
learning low-dimensional yet representations of raw (or minimally processed)
data. Additionally, NPE is more robust than ABC to a less-than-optimal proposal
distribution (e.g. a less-informative or miscalibrated prior), since ABC
requires that many simulations match the observed data closely. For these
reasons, we argue that NPE may scale better than ABC for a range of
high-dimensional datasets and mechanistic models found in epidemiology and
related fields. Training neural networks can be costly in terms of computational
resources and engineering labor, yet we obtained our results with a minimalist
NPE ``stack'' of a feedforward encoder architecture and a Gaussian variational
approximation, highlighting a methodology for rapid, approximate inference.

When calibrating our heterogeneous transmission model to the CRKP data, we found
potential evidence of model misspecification. Though NPE achieved strong
empirical accuracy (Table~\ref{tab:mspe}), it failed to reproduce the
likelihood-based MCMC parameter estimate, unlike in the simulation experiment. The validity of
NPE depends on the simulator being a reasonable description of the real-world
data generating process, that is, a well-specified model. Thus, neural SBI requires an even
greater degree of trust in model design than classical statistical inference.
For noisy real-world datasets, designing a faithful individual-level model of
transmission may be unworkable. Judicious use of summary statistics may help
mitigate the worst effects of misspecification (\cite{frazier2021robust,
ward2022robust}), but these risk discarding relevant information in the data. 

We have put forward NPE as a viable alternative to methods such as MCMC and ABC
for calibrating complex mechanistic models, but these have better understood
statistical properties in comparison, such as asymptotic
convergence.\footnote{For instance, under a misspecified model, likelihood-based
Bayesian inference will yield posteriors that concentrate on the ``pseudotrue''
parameters that minimize the KL divergence between the true model and the
misspecified model (\cite{frazier2020model}).} Nevertheless, using NPE to find Gaussian
posterior approximations may offer some theoretical grounding. The Bernstein-von
Mises theorem suggests that the true posterior for global model parameters,
under basic regularity conditions, will approximate a multivariate normal
distribution as the sample size increases. \citeauthor{mcnamara2024globally}
showed that NPE, when using a single-layer encoder architecture and an
exponential variational family, will converge on the \textit{global} optimizer
for the NPE objective function (Equation~\ref{eq:npe-obj}). Together, these
results hint at the possibility of explaining NPE's output, although more theoretical
work in this direction remains.

We focused on an application to an epidemiological dataset where infection is
effectively fully observed, making likelihood-based inference feasible, so as to
critically examine simulation-based inference in the real world. The practical use
case for NPE in epidemiology is modeling systems with stochastic events that are
both heterogeneous and latent (i.e. unobserved), thereby yielding intractable
likelihoods. When models are complex enough that simulation is
computationally slow, NPE may show efficiency advantages over ABC. NPE shows
promise in connecting deep learning's empirical predictive power to
interpretable, mechanistic models of nature, but this synthesis hinges on
careful model design grounded in data and domain expertise.

\subsection{Future Work}\label{sec:future}

As the cost of whole-genome sequencing (WGS) of pathogens has decreased, it has
become commonplace to integrate genomic data into epidemiological analysis to
identify plausible individual-level transmission pathways
(\cite{kao2014supersize, klinkenberg2017simultaneous}). As part of the
intervention against CRKP carried out in the Chicago LTACH, WGS was conducted
for virtually all CRKP isolates (\cite{hawken2022threshold}), making it feasible
to infer the disease's phylogeny as it spread through the patient population. In future work, we
plan on developing a \textit{phylodynamic} simulation model for CRKP
transmission that combines epidemiological time series data with phylogenic
data. This simulator may give insight into transmission risks at a more granular
level and would likely benefit from the scalability advantages offered by NPE.
To better understand the epidemiology of HAIs and design
effective prevention strategies, our mechanistic model could be extended to
describe the spread of disease across multiple healthcare facilities in a region
(\cite{han2019whole}), since HAIs are often imported into facilities from the
outside.

In this work, we have investigated stochastic epidemiological modeling in the
context of HAIs, which are relatively closed and small-scale systems, yet our
methodology could be applied to a broad range of epidemiological problems. Many
infectious diseases are best described by stochastic compartmental models with
unobserved states, such as the classic Susceptible-Infected-Recovered (SIR)
model: NPE could be useful for fitting SIR-type simulators without needing to
recover computationally-expensive complete likelihoods. Time series data feature
prominently in epidemiology, suggesting the use of sequential neural networks
architectures like Recurrent Neural Networks and Transformers to learn
time-varying transmission patterns (cf. \cite{madden2024deep}). We have focused
on parameter estimation with NPE for a given simulation, but this is only the
first stage of principled Bayesian inference, which in its entirety should
involve the criticism of and selection among multiple models
(\cite{mackay1992bayesian}). This is a hard problem in the absence of explicit
likelihoods (\cite{spurio2023bayesian}). Further research into criticism of
simulation models and diagnosing misspecification (\cite{ward2022robust}) will
contribute greatly to the utility of NPE for applied data analysis.

\section{Significance Statement}\label{sec:signif} 

Realistic simulation models of infectious disease transmission often lack
computationally tractable likelihoods. Thus, fitting these models to real-world
data using classical, likelihood-based statistical tools (e.g. MCMC) can be
impractical. In this work, we train neural networks on simulated data generated
from epidemiological models to infer plausible values of unknown parameters
corresponding to observed data. Assuming that the given simulation model is a
reasonable description of the real-world epidemic system, this approach, NPE,
can achieve accurate calibration with scalability advantages compared to the
well-established technique of matching simulations against observed data (ABC).
Our methodology can potentially help epidemiologists innovate and deploy complex
simulation models in a variety of applied problems.

\begin{acks}[Acknowledgments]
The authors thank Declan McNamara for conceptual advice on NPE, Hannah Steinberg
for help interpreting the Chicago CRKP data, and Krzysztof Sakrejda for feedback
on the presentation of results.
\end{acks}
\begin{funding}
PC was supported by a Propelling Original Data Science (PODS) grant from the
Michigan Institute for Data and AI in Society (MIDAS). Our use of the data was
made possible by funding from the Centers for Disease Control and Prevention
(CDC 5U54CK000607). ES was supported by NIH award 5R01AI175227.
%
\end{funding}

\section*{Software}
Open-source code with our implementation of NPE,
our simulation models, and experiments is available at
\url{https://github.com/epibayes/np-epid}.

\bibliographystyle{imsart-nameyear} 
\bibliography{main.bib}       

\begin{appendix}

\section{Susceptible-Infected Model}

\subsection{SI Model Likelihood Derivation}
\label{supp:si-derivation}

We derive the likelihood written out in equations 12, 13, and 14 from the following
individual transition probabilities. For $1 \leq t \leq T,$
\begin{itemize}
\item $P(X^{(i)}_t = 0 \mid X^{(i)}_{t-1} = 0) = \gamma (1 - \alpha) + (1 -
\gamma)(e^{-\lambda_i(t)})$
\begin{itemize}
    \item A susceptible patient is discharged and replaced with a susceptible or
    is not discharged and fails to get infected
\end{itemize}
\item $P(X^{(i)}_t = 0 \mid X^{(i)}_{t-1} = 1) = \gamma(1 - \alpha)$ 
\begin{itemize}
    \item An infected patient is discharged and replaced with a susceptible
\end{itemize}
\item $P(X^{(i)}_t = 1 \mid X^{(i)}_{t-1} = 0) = \gamma \alpha + (1 - \gamma)(1
- e^{-\lambda_i(t)})$
\begin{itemize}
    \item A susceptible patient is discharged and replaced with an infected or
    is not discharged and gets infected
\end{itemize}

\item $P(X^{(i)}_t = 1 \mid X^{(i)}_{t-1} = 1) = \gamma \alpha + (1 - \gamma)$
\begin{itemize}
    \item An infected patient is discharged and replaced with an infected or is
    not discharged
\end{itemize}
\end{itemize}
The asymmetry in transition probabilities between already-infected and
already-susceptible patients arises from the assumption that an infected patient
stays infected without recovery. It follows that the transition likelihood for
an individual patient is
\begin{equation}
\begin{split}
P(X_t^{(i)} \mid X_{t-1}^{(i)}) = \left [ \gamma \alpha + (1 - \gamma) (1 - e^{-\lambda_i(t) })^{(1 - X^{(i)}_{t-1})} \right ]^{X^{(i)}_t} \\
\cdot \left [ \gamma (1 - \alpha) + (1 - \gamma) (e^{- \lambda_i(t) } ) (1 - X^{(i)}_{t-1} )\right ] ^ {(1 - X^{(i)}_{t} )}.
\end{split}
\end{equation}

\subsection{SI Model Algorithm}
\label{supp:si-algorithm}

\begin{algorithm}[H]
    \caption{Stochastic Discrete-time SI Simulator}\label{alg:si}
    \begin{algorithmic}
        \Require Vector of transmission rates $\boldsymbol{\beta}$, discharge
        probability $\gamma$, population proportion of infected $\alpha$, floor
        assignments $\mathbf{k}$, room assignments $\mathbf{r}$ \Ensure $N
        \times T$ matrix $\mathbf{X}$ of infection logs $\mathbf{X}^{(i)} =
        \{X^{(i)}_1 \ldots X^{(i)}_T \}$ for all patients $i=1, \ldots, N$
        \State Initialize infecteds as $X^{(i)}_1 \sim \text{Bernoulli}(\alpha)$
        \For{$t = 2, \ldots T$} \For{$i \in 1, \ldots, N$} \State Draw
        $D_t^{(i)} \sim \text{Bernoulli}(\gamma)$ \If{$D_t^{(i)} = 1$}
        \Comment{patient $i$ is discharged and replaced} \State Draw $X_t^{(i)}
        \sim \text{Bernoulli}(\alpha)$ \Else \If{$X^{(i)}_{t-1} = 0$}
        \Comment{patient $i$ is susceptible} \State Compute the individualized
        force of infection $\lambda_i(t)$ \State Draw $X_t^{(i)} \sim
        \mbox{Bernoulli}(1 - e^{-\lambda_i(t)})$ \Else \State $X_t^{(i)} \gets
        1$ \EndIf \EndIf \EndFor \EndFor
    \end{algorithmic}
\end{algorithm}

\subsection{Summary Statistics}
\label{supp:summ-stats}

To estimate the vector of heterogeneous infection rates $\bbeta,$ we process the
raw infection status data $\bX$ into seven location-specific summary statistics
or ``views,'' which we write as $\bJ.$ We can think of $\bJ$ as a $T \times
(K+2)$ matrix, where $K$ is the number of floors. Written in column form, 
\begin{equation}
    \bJ = \begin{pmatrix} \bI & \mathbf{L}_1 & \ldots & \mathbf{L}_K & \mathbf{R} \end{pmatrix},
\end{equation}
where
\begin{itemize}
    \item $\bI = \{I_1, \ldots, I_T \}$ is the overall case count over time.
    Recall that $I_t = \sum_{i = 1}^N X_t^{(i)}.$ This is the same summary
    statistic used for estimating the homogeneous infection rate.
    \item For every $k = 1, \ldots, K$ and $t = 1, \ldots, T,$ define
    \begin{equation}
        L_{k, t} = \sum_{i: F(i) = k} X_t^{(i)},
    \end{equation}
    with $F(i)$ indicating the floor on which patient $i$ resides. Then,
    $\mathbf{L}_k = \{L_{k, 1}, \ldots, L_{k, T} \}.$ This is the number of
    cases over time on each floor. 
    \item For each $t = 1, \ldots, T$ and $i = 1, \ldots, N,$ we define
    $Q_t^{(i)},$ to be the number of infected patients at time $t$ in the room
    belonging to patient $i$. Suppose there are $N_R$ rooms.  We define
    \begin{equation}
        R_t = \# \{i: \ Q_t^{(i)} > 1 \}
    \end{equation}
    
Then, $\mathbf{R} = \{R_1, \ldots, R_T \}.$ We can think of this statistic as
measuring the number of rooms with multiple infected patients, which should
intuitively correlate with the risk of roommate-roommate transmission.
\end{itemize}

It is greatly beneficial for deep learning models to standardize input data such
that variance is comparable across multiple features. To this end, we rescale
all statistics so that they fall between 0 and 1. We divide $I$ by $N$, the
overall patient population. When the facility population variable, as in the
CRKP dataset, we divide by the maximum observed population over the period of
study. Likewise, we divide $\mathbf{L}_1, \ldots, \mathbf{L}_K$ by the
population (or maximum capacity) of each floor, and divide $\mathbf{R}$ by the
number of unique rooms.

\subsection{CRKP Transmission Model Likelihood}
\label{supp:crkp-model}

Let $\mathbf{W}$ denote the $N \times T$ \textit{facility trace} matrix. For any patient $i$,
$W_t^{(i)}$ equals 1 if patient $i$ is present in the facility during week $t$ and zero otherwise. The model
likelihood is then

\begin{equation}
\log \mathcal{L}(\beta ; \mathbf{X}) = \sum_{t=2}^T \sum_{i=1}^N [(1 - X_{t-1}^{(i)})W_t^{(i)}W_{t-1}^{(i)}] \cdot [X_t^{(i)} \log (1 - e^{- \lambda_i(t)}) + (1 - X_t^{(i)}) \log (e^{- \lambda_i(t)})],
\end{equation}
with the individualized force of infection being computed as a function of $\bbeta$ according to Equations 6 and 7.

At time step 1, we don't simulate any infections. Let $2 \leq t \leq T$. For any present patient $i$ ($W_t^{(i)} = 1$), we consider the following cases:

\begin{enumerate}
    \item Patient $i$ has just been admitted to the facility ($W_t^{(i)} = 1 \wedge W_{t-1}^{(i)} = 0$): the patient's status $X_t^{(i)}$ is determined by their screening results, no contribution to the likelihood
    \item Patient $i$ is already present in the facility and was infected at time $t-1$  ($W_t^{(i)} = 1 \wedge W_{t-1}^{(i)}  = 0 \wedge X_{t-1}^{(i)} = 1 $): this patient stays infected ($ X_{t}^{(i)} = 1 $)
    \item Patient $i$ is already present in the facility and was susceptible at time $t-1$ ($W_t^{(i)} = 1 \wedge W_{t-1}^{(i)}  = 0 \wedge X_{t-1}^{(i)} = 0 $): we randomly sample this patient's current infection status,
    $$
    X_{t}^{(i)} \sim \text{Bernoulli}(1 - e^{-\lambda_i(t)})
    $$
\end{enumerate}

Hence, 
\begin{equation}
P(X_t | X_{t-1}) = \prod_{i=1}^N \left [ (1 - e^{- \lambda_i(t)}) ^ { X_{t}^{(i)} } (e^{- \lambda_i(t)}) ^ { 1 - X_{t}^{(i)} } \right ] ^ { (1 - X_{t-1}^{(i)}) W_t^{(i)} W_{t-1}^{(i)}}
\end{equation}
and the complete likelihood is given by

\begin{equation}
    P(\bX) = \prod_{t=2}^T P(X_t | X_{t-1}).
\end{equation}

\section{Partial Observation Model}
\label{supp:partial}
\subsection{Model Formulation}

A patient colonized with a nosocomial disease will often show no symptoms unless
the infection becomes invasive (e.g., bacteria move from the surface of the skin
to internal organs). For this reason, it is unrealistic to assume that all
patient infection statuses will be observed, barring extensive prospective
surveillance of individuals: a majority of infected patients may be asymptomatic
carriers. To address this problem, we extend the stochastic SI model to the
scenario of \textit{partial observation}.

We introduce a new status variable $Y_t^{(i)},$ that indicates whether the
patient at location $i$ is observably infected by time step $t.$ Analogously to
$X_t^{(i)},$ let $Y_t = (Y_t^{(1)}, \ldots, Y_t^{(N)})$ and let $\bY = Y_1,
\ldots, Y_T$: $\bY$ denotes our observed data, the observed cases over time. In
this variant of the SI model, the old status variable $X_t^{(i)}$ reflects
asymptomatic infection (colonization) with the pathogen of interest. We observe
a colonized patient's infection either because they are experiencing invasive
symptoms or because they were screened for pathogens upon entry. We extend our
data generating process to $Y$ on the basis of four simple rules.
\begin{enumerate}
    \item If a location $i$ admits a new patient at time step $t,$ they are
    tested upon admission, then we observe their status. Therefore, $Y_t^{(i)} =
    X_t^{(i)}.$
    \item At any time step $t,$ a colonized but hitherto asymptomatic patient
    will show symptoms with probability $\eta \in (0, 1]$. That is, $Y_t^{(i)}
    \sim \text{Bernoulli}(\eta)$.
    \item An uncolonized patient will never show symptoms: $X_t^{(i)} = 0
    \implies  Y_t^{(i)} = 0$
    \item A symptomatic patient will remain symptomatic. Assuming there is no
    turnover at location $i$ during time step $t,$ $Y_{t-1}^{(i)} = 1 \implies
    Y_{t}^{(i)} = 1.$
\end{enumerate}
$\eta$ is the \textit{probability of observation}, which we can interpret as the
chance of an asymptomatic infection going invasive during any one time step.
While we might have reason to believe that the probability of observation
depends on individual-level covariates (e.g. age, comorbidities) and/or time
spent in the facility, for simplicity's sake we assume a homogeneous and
constant $\eta.$ In this model, the time it takes for a colonized patient to
show symptoms follows a geometric distribution, which can be thought of as a
discretized exponential distribution. The formula for the force of infection is
the same as for the fully observed heterogeneous SI model. 
In our simulation experiments, we treat $\eta$ as a known parameter,
which may be an unrealistic assumption. However, it is difficult to identify the observation 
probability---that is, the propensity
toward missingness---of cases from case data alone.

Below, we outline the simulation program corresponding to the forward data
generating model under partial observation of cases. With some effort, we could
write out the \textit{complete likelihood}, $p(\bY, \bX \mid \bbeta) = p(\bY
\mid \bX) p(\bX \mid \bbeta)$, using the rules above to compute the conditional
probability of observed infection $p(\bY \mid \bX)$. Computing the observed data
likelihood $p(\bY \mid \bbeta)$ is infeasible at scale, since it would entail
integrating the complete likelihood over all possible configurations of $\bX$.
This is a massive state space when the force of infection is heterogeneous.
Heuristically speaking,
\begin{equation}
    p(\bY \mid \bbeta) \approx \sum_{\bX} p(\bY, \bX \mid \bbeta).
\end{equation}
$\bX$ has $2^{T \cdot N}$ configurations, though not all of these
are permissible given a set of observed infection times $\bY$. Therefore, the
observed data likelihood has a (worst-case) complexity of $O(2^{T \cdot N} T
\cdot N^2)$.

\begin{algorithm}
\caption{Stochastic SI Simulator with Partial Observation}\label{alg:si-partial}
\begin{algorithmic}
\Require Vector of transmission rates $\boldsymbol{\beta}$, discharge
probability $\gamma$, population proportion of infected $\alpha$, floor
assignments $\mathbf{k}$, room assignments $\mathbf{r},$ probability of
observation $\eta$ \Ensure $N \times T$ matrix $\mathbf{Y}$ of observed case
logs $\mathbf{Y}^{(i)} = \{Y^{(i)}_1 \ldots Y^{(i)}_T \}$ for all patients $i=1,
\ldots, N$ \State Initialize array of colonization statuses $X^{(i)}_1 \sim
\text{Bernoulli}(\alpha)$ \State Set $Y^{(i)}_1 \gets X^{(i)}_1$ \For{$t = 2,
\ldots T$} \For{$i \in 1, \ldots, N$} \State Draw $D_t^{(i)} \sim
\text{Bernoulli}(\gamma)$ \If{$D_t^{(i)} = 1$} \Comment{patient $i$ is
discharged and replaced} \State Draw $X_t^{(i)} \sim \text{Bernoulli}(\alpha)$
\Else \If{$X^{(i)}_{t-1} = 0$} \Comment{patient $i$ is susceptible} \State
Compute the individualized force of infection $\lambda_i(t)$ \State Draw
$X_t^{(i)} \sim \mbox{Bernoulli}(1 - e^{-\lambda_i(t)})$ \Else \State $X_t^{(i)}
\gets 1$ \EndIf \EndIf \If{$D_t^{(i)} = 1$} \Comment screen newly admitted
patient $i$ \State $Y_t^{(i)} \gets X_t^{(i)}$ \ElsIf{$X_t^{(i)} = 1 \wedge
Y_{t-1}^{(i)} = 0$} \State Draw $Y_t^{(i)} \sim \mbox{Bernoulli}(\eta)$ \Else
\State $Y_t^{(i)} \gets Y_{t-1}^{(i)}$ \EndIf \EndFor \EndFor
\end{algorithmic}
\end{algorithm}

\subsubsection{Complete Likelihood}

We sketch a derivation of $p(\bY \mid \bX)$ in the simplest case where $\gamma =
0$, that is, there is no random turnover of patients.
$$
P(Y_1^{(i)} \mid X_1^{(i)}) = \begin{cases}
    (1 - \eta)^{X_1^{(i)}} &{\text{if }}Y_1^{(i)} = 0\\
    1 - (1 - \eta)^{X_1^{(i)}} &{\text{if }}Y_1^{(i)} = 1
\end{cases}
$$
and for $t > 1$,
$$
P(Y_t^{(i)} \mid Y_{t-1}^{(i)}, X_1^{(i)}, \ldots,  X_t^{(i)}) = \begin{cases}
    (1 - \eta)^{W_t^{(i)}} & \text{if }Y_t^{(i)} = 0 \wedge Y_{t-1}^{(i)} = 0\\
    0 & \text{if }Y_t^{(i)} = 0 \wedge  Y_{t-1}^{(i)} = 1\\
    1 - (1 - \eta)^{W_t^{(i)}} & \text{if }Y_t^{(i)} = 1 \wedge Y_{t-1}^{(i)} = 0\\
    1 & \text{if }Y_t^{(i)} = 1 \wedge Y_{t-1}^{(i)} = 1
\end{cases},
$$
where $W_t^{(i)} = \sum_{s=1}^t X_s^{(i)}$. Let $S_i = \min \{t: Y_t^{(i)} = 1
\}$ (or $T+1$ if this is undefined). Let $\tilde S_i = \min \{t: X_t^{(i)} = 1
\}$ (or $T+1$ if this is undefined).
$$
P(Y_t^{(i)}  = 1 \mid X_1^{(i)}, \ldots,  X_t^{(i)} ) = P(S_i \leq t \mid  X_1^{(i)}, \ldots,  X_t^{(i)} ),
$$ 
and $S_i - \tilde S_i$ follows a geometric distribution. Thus,
$$
P(S_i \leq t \mid  X_1^{(i)}, \ldots,  X_t^{(i)} ) = P(S_i - \tilde S_i \leq t  - \tilde S_i \mid  X_1^{(i)}, \ldots,  X_t^{(i)} )
$$
$$
= 1 - (1 - \eta)^{t - \tilde S_i + 1} = 1 - (1 - \eta)^{W_t^{(i)}}.
$$ 
Therefore we can compute
$$
P(Y_i \mid X_i) = P(Y_1^{(i)} \mid X_1^{(i)}) \prod_{t=2}^T P(Y_t^{(i)} \mid Y_{t-1}^{(i)}, X_1^{(i)}, \ldots,  X_t^{(i)}) 
$$
and 
$$
P(\bY \mid \bX) = \prod_{i=1}^N P(Y_i \mid X_i)
$$

Finally,
$$
p(\bY \mid \bbeta) =  \int p(\bY, \bX \mid \bbeta) d\bX,
$$
where 
$$
p(\bY, \bX \mid \bbeta) = p(\bY \mid \bX) p(\bX \mid \bbeta),
$$
where $p(\bX \mid \bbeta)$ is the likelihood under complete observation.

\subsection{Simulation Experiments}

For our last simulated experiment, we turn to the problem of calibrating a
stochastic model to a partially-observed epidemic. Incomplete observation of
infection times results in a model with a highly intractable likelihood function
due to the large number of latent variables. However, forward simulation from
such models is simple, and our previous experiments have shown that NPE
accurately estimates model parameters from data simulation alone. With this in
mind, we modify the heterogeneous simulation setup (Section 3.3) so that
$\eta = 0.1$: that is, an expected 10\% patients with asymptomatic colonization
show observable symptoms of infection.

In Figure 1a, we compare the observed case count over time to
the actual number of infected patients. The former includes individuals who are
screened upon arrival and colonized individuals who develop symptoms (with
probability $\eta$); the latter comprises both observed cases and asymptomatic
but infectious carriers. By the end of the period of study, approximately 20\%
of patients in the facility are asymptomatically colonized. The majority of
infectious are observed throughout, despite $\eta$ being close to zero. We show
a breakdown of observed cases by floor with the facility-level case rate as a
comparison in Figure 1b. One effect of partial
observation is that the location-based heterogeneity in transmission is masked
by a greater level of noise. For example, while Floor 3 has the third highest
infection rate (see Table 1 for exact parameter values), for
much of the period of observation, Floor 3 exhibits the highest infection
burden.

\begin{figure}
    \centering
    \begin{subfigure}[t]{0.45\textwidth}
        \centering
        \includegraphics[width=\textwidth]{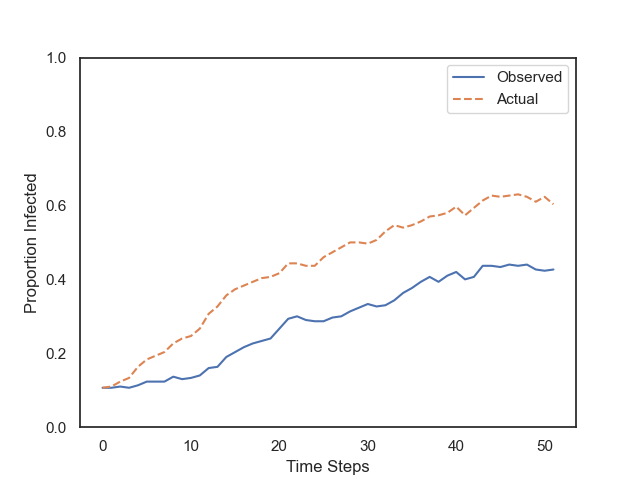}
        \caption{Comparison of the observed (symptomatic) and actual proportion of infected patients over time with $\eta = 0.1$}
        \label{fig:partial_obs}
    \end{subfigure}
    \hfill
    \begin{subfigure}[t]{0.45\textwidth}
        \centering
        \includegraphics[width=\textwidth]{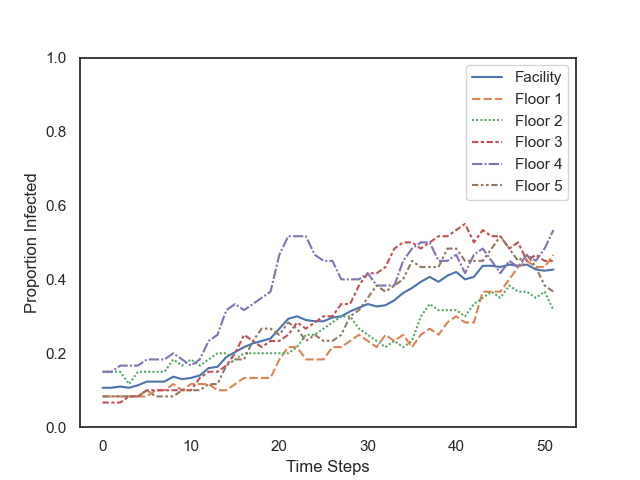}
        \caption{Observed proportion of infected over time for the facility and by floor}
        \label{fig:partial_floor_viz}
    \end{subfigure}
    \caption{Visualization of a simulated epidemic realization under an SI model with partial observation and heterogeneous transmission.}
    \label{fig:partial}
\end{figure}

To estimate the unknown infection rates, we fit ABC and NPE with a MVN posterior
approximation to simulated data, using the same diagonal lognormal prior as in
Section 4.3
We used 4,000 samples to train NPE and took 100 ABC samples from 11,779
simulations. In Tables 1 and 2, we compare the two methods' posterior mean point estimates
and marginal standard deviations (on the log scale) for each infection rate.

In Table 1, we compare the two methods' posterior mean point estimates and 90\%
(corresponding to the (0.05, 0.95) quantile range). The NPE mean estimates are
overall close to the ABC estimates, despite NPE having been fit with less than
one third of the samples. Both methods differentiate infection rates by floor,
with the ordering of rate sizes more or less reflecting the observed incidence
by floor (Figure 1b). While unbiasedness is not a property of Bayesian
estimators, the NPE estimates appear to be closer to the true parameter values
than the ABC estimates. ABC exhibits lower posterior uncertainty than NPE (Table
2), though NPE seems to identify the presumed negative pairwise correlation
between the facility rate $\beta_0$ and location-specific rates better than ABC.

\begin{table}
\caption{Posterior mean estimates of infection rates in a partially-observed outbreak.}
\centering
\begin{tabular}{llll}
\toprule
Transmission Rate & Value & NPE Mean & ABC Mean\\
\midrule
Facility & 0.05 & 0.0820 & 0.0919 \\
Floor 1 & 0.02 & 0.03051 & 0.0183 \\
Floor 2 & 0.04 & 0.0215 & 0.0187 \\
Floor 3 & 0.06 & 0.0563 & 0.0341 \\
Floor 4 & 0.08 & 0.0624 & 0.0452 \\
Floor 5 & 0.1 & 0.0396 & 0.0243 \\
Room6 & 0.05 & 0.0550 & 0.0299 \\
\bottomrule
\end{tabular}
\label{tab:partial}
\end{table}

\begin{table}
\caption{Marginal standard deviation estimates of infection rates (log-scale) in a partially-observed outbreak.}
\centering
\begin{tabular}{llll}
\toprule
Transmission Rate & Value & NPE Mean & ABC Mean\\
\midrule
Facility & -3.00 & 0.383 & 0.244 \\
Floor 1 & -3.91 & 0.924 & 0.877 \\
Floor 2 & -3.22 & 1.05 & 0.803 \\
Floor 3 & -2.81 & 0.837 & 0.948 \\
Floor 4 & -2.53 & 0.847 & 1.00 \\
Floor 5 & -2.30 & 0.812 & 0.796 \\
Facility & -3.00 & 1.04 & 0.986 \\
\bottomrule
\end{tabular}
\end{table}

In the absence of a likelihood-based estimate of the infection rates, posterior
predictive checks offer some sense of the reliability of simulation-based
estimates. In Figure 2, we visualize posterior predictive checks for all spatial
descriptive statistics in $\bJ$. The NPE posterior predictive distribution shows
substantial volatility, though this could reflect the inherent aleatoric
uncertainty of partial observation of cases. NPE appears to calibrate well to
most locations, though it appears to overestimate the incidence on Floor 1. Overall, the NPE
posterior predictions were similar to the ABC predictions, including similar patterns of
bias, though the latter had less variance.

\begin{figure}
\begin{subfigure}[t]{.4\textwidth}
\centering
\includegraphics[width=\linewidth]{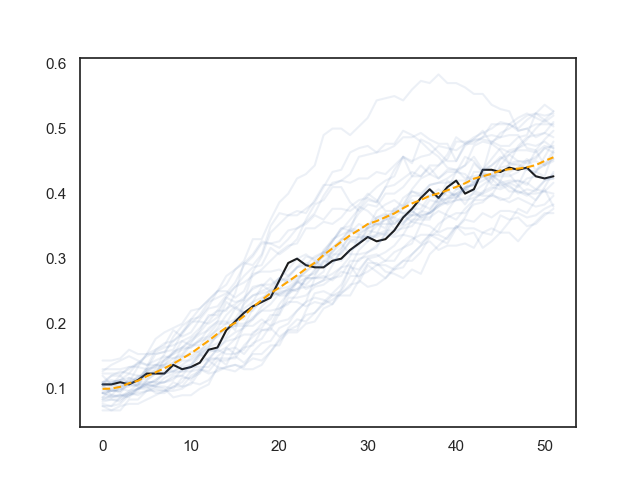}
\caption{Facility}
\label{}
\end{subfigure}
\hfill
\begin{subfigure}[t]{.4\textwidth}
\centering
\includegraphics[width=\linewidth]{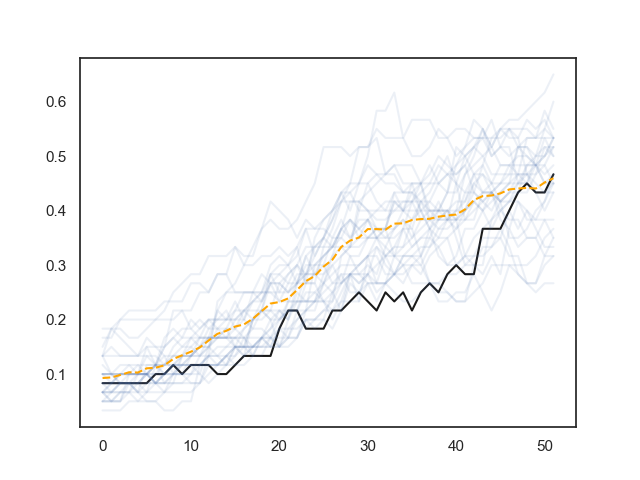}
\caption{Floor 1}
\label{}
\end{subfigure}

\medskip

\begin{subfigure}[t]{.4\textwidth}
\centering
\includegraphics[width=\linewidth]{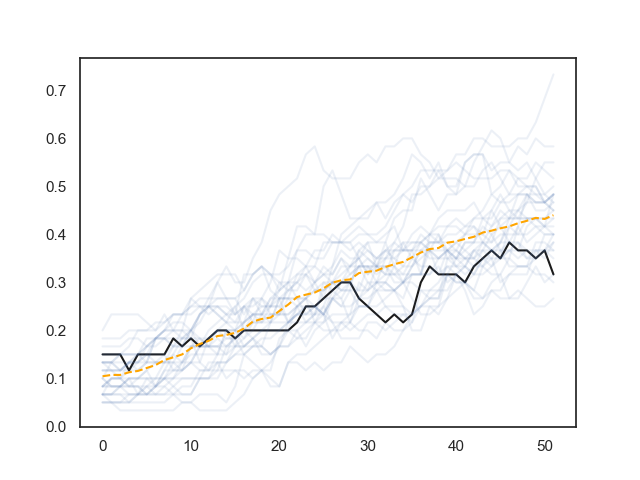}
\caption{Floor 2}
\label{}
\end{subfigure}
\hfill
\begin{subfigure}[t]{.4\textwidth}
\centering
\includegraphics[width=\linewidth]{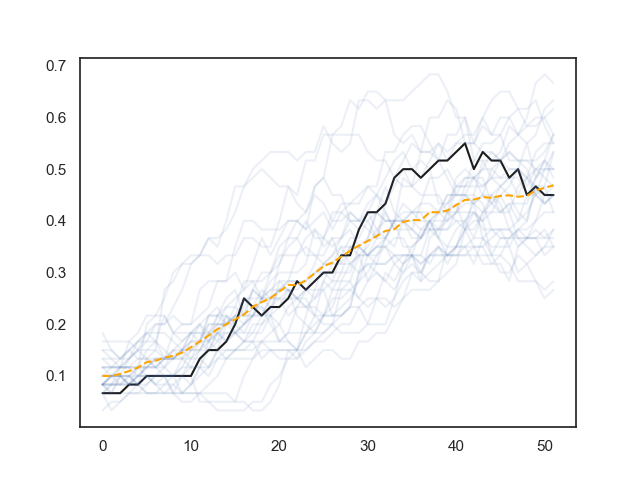}
\caption{Floor 3}
\label{}
\end{subfigure}

\medskip

\begin{subfigure}[t]{.4\textwidth}
\centering
\includegraphics[width=\linewidth]{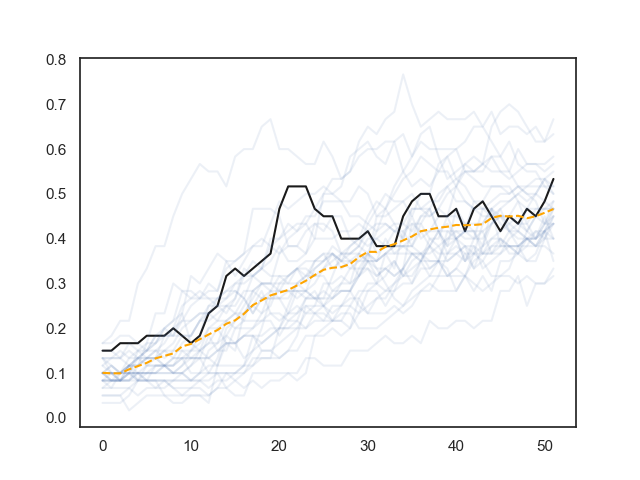}
\caption{Floor 4}
\label{}
\end{subfigure}
\hfill
\begin{subfigure}[t]{.4\textwidth}
\centering
\includegraphics[width=\linewidth]{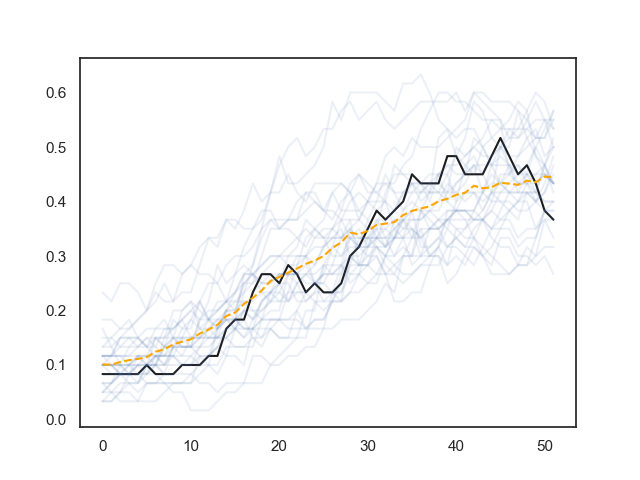}
\caption{SCU (Floor 5)}
\label{}
\end{subfigure}

\par

\centering
\begin{subfigure}[t]{.4\textwidth}
\centering
\includegraphics[width=\linewidth]{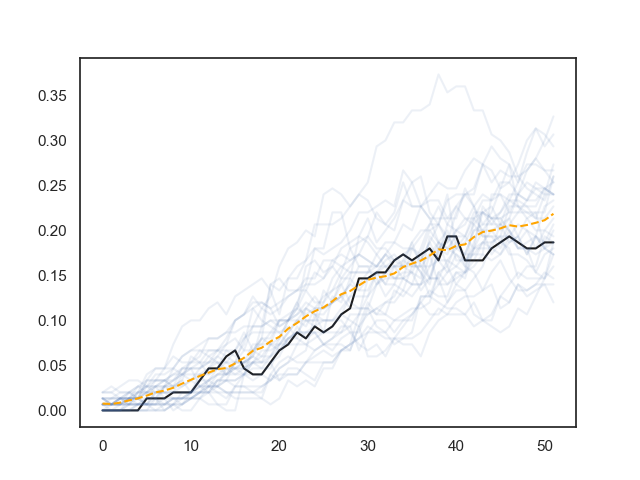}
\caption{Room}
\label{}
\end{subfigure}

\caption{Predictive checks of the NPE posterior estimate.}
\end{figure}

\begin{figure}
    \centering
    \begin{subfigure}[t]{0.45\textwidth}
        \centering
        \includegraphics[width=\textwidth]{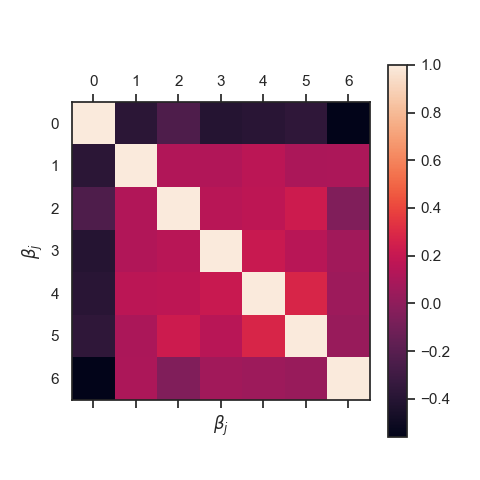}
        \caption{NPE posterior correlations (truncated MVN approximation)}
        \label{fig:partial-corr-a}
    \end{subfigure}
    \hfill
    \begin{subfigure}[t]{0.45\textwidth}
        \centering
        \includegraphics[width=\textwidth]{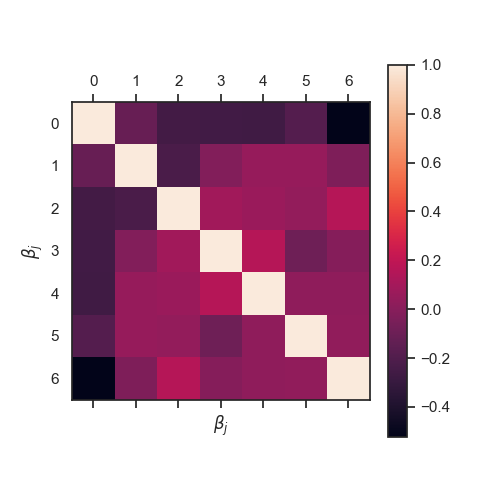}
        \caption{ABC posterior correlations}
        \label{fig:partial-corr-b}
    \end{subfigure}
    \caption{Correlation heatmaps for the simulation-based approximate posterior distribution of heterogeneous infection rates $\bbeta$ under partial observation of cases.}
    \label{fig:partial-corr}
\end{figure}

\section{Supplemental Figures and Tables}

\subsection{Simulation Experiments}
\label{supp:sim-figures}


\begin{figure}
    \centering
    \includegraphics[width=0.5\linewidth]{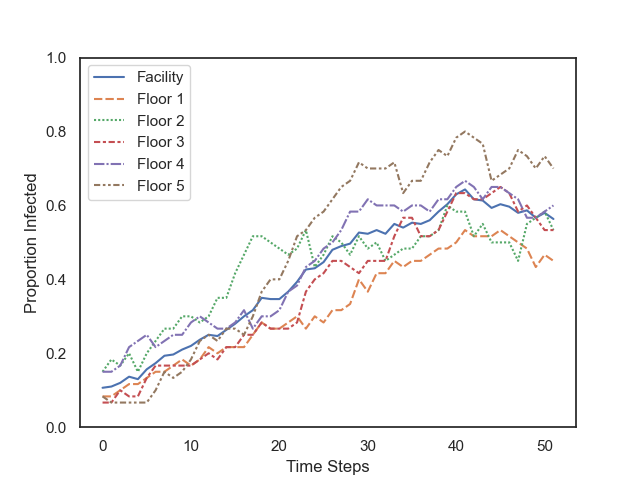}
    \caption{Facility and floor-level incidence.}
\end{figure}

\begin{figure}
\centering
\includegraphics[width=0.6\linewidth]{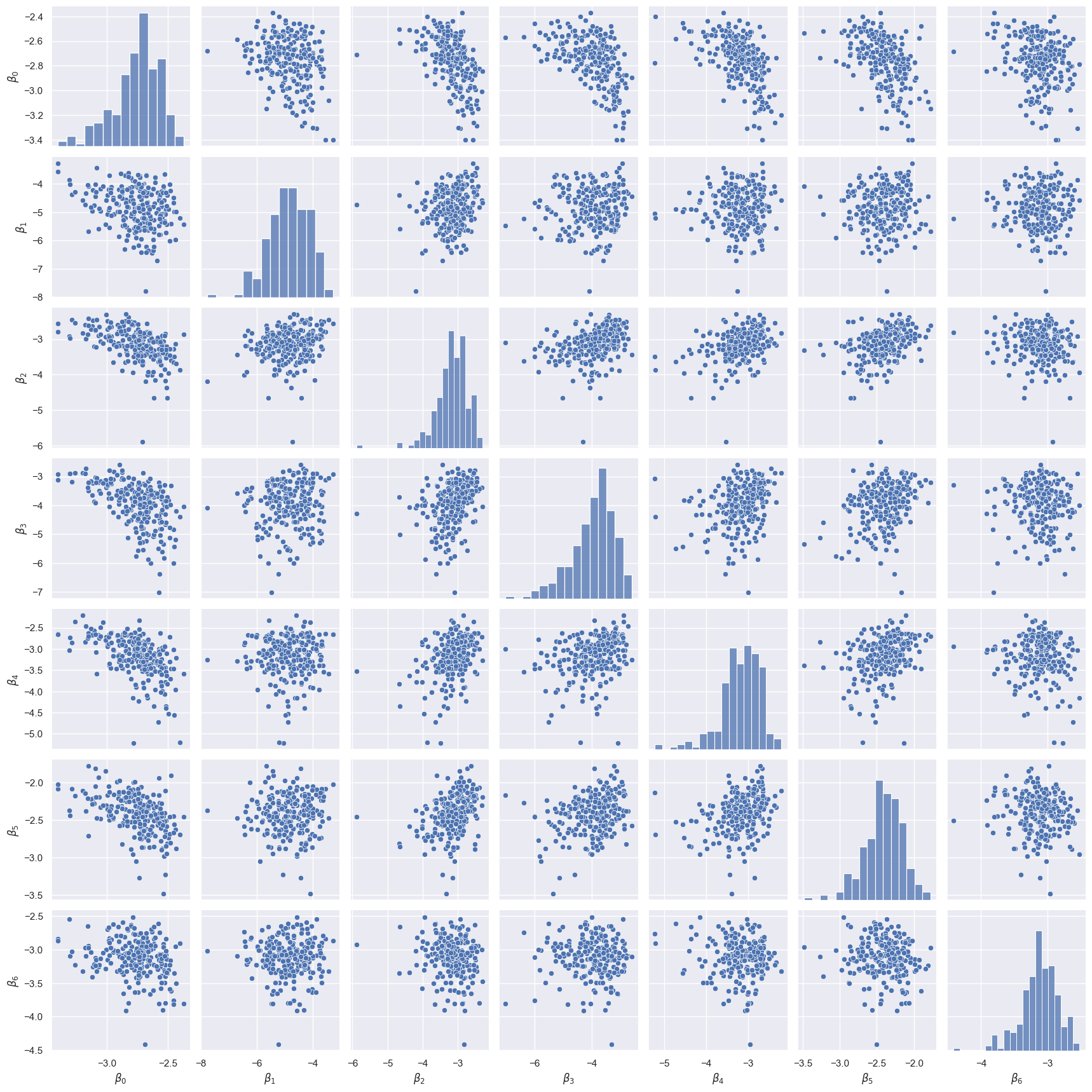}
\caption{Pair plot of the exact, likelihood-based sample of heterogeneous infection rates on the log scale. Several of
the covariances between $\beta_0$ and other components appear non-elliptical and thus non-Gaussian.}
\label{fig:si-pairs}
\end{figure}

\begin{figure}
\begin{subfigure}[t]{.45\textwidth}
\centering
\includegraphics[width=\linewidth]{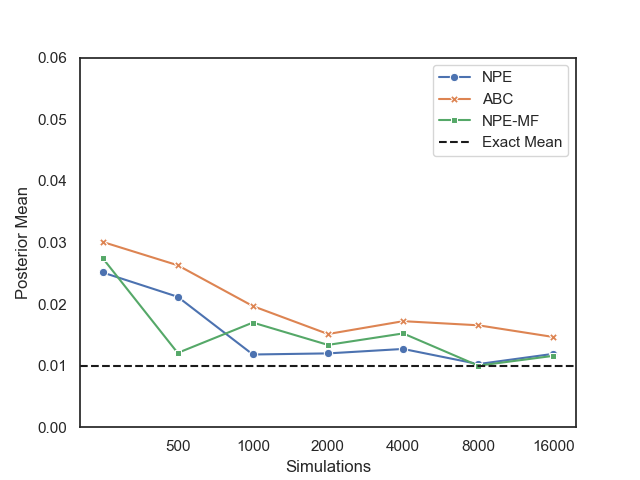}
\caption{Estimation of the infection rate within Floor 1, $\beta_1$.}
\label{}
\end{subfigure}
\hfill
\begin{subfigure}[t]{.45\textwidth}
\centering
\includegraphics[width=\linewidth]{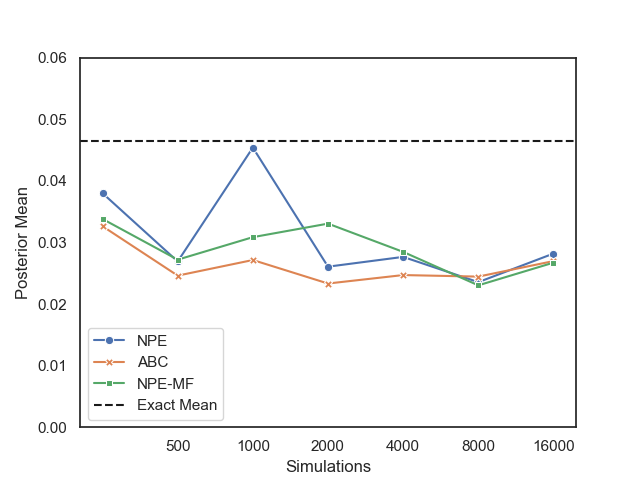}
\caption{Estimation of the infection rate within Floor 2, $\beta_2$.}
\label{}
\end{subfigure}

\medskip

\begin{subfigure}[t]{.45\textwidth}
\centering
\includegraphics[width=\linewidth]{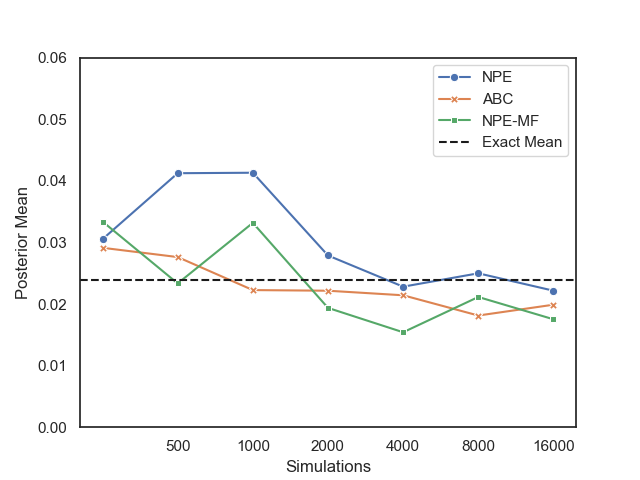}
\caption{Estimation of the infection rate within Floor 3, $\beta_3$.}
\label{}
\end{subfigure}
\hfill
\begin{subfigure}[t]{.45\textwidth}
\centering
\includegraphics[width=\linewidth]{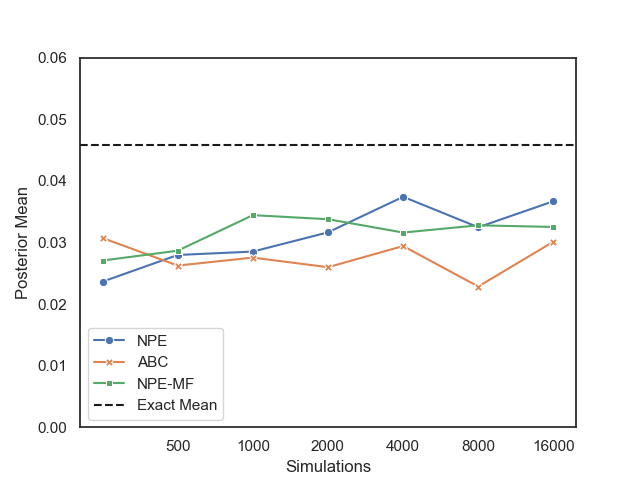}
\caption{Estimation of the infection rate within Floor 4, $\beta_4$.}
\label{}
\end{subfigure}
\caption{Simulation-based estimation accuracy and sample-efficiency for heterogeneous infection rates (simulated experiment).}
\label{fig:het-error-complete}
\end{figure}

\begin{figure}
\centering
\begin{subfigure}[t]{0.45\textwidth}
    \centering
    \includegraphics[width=\textwidth]{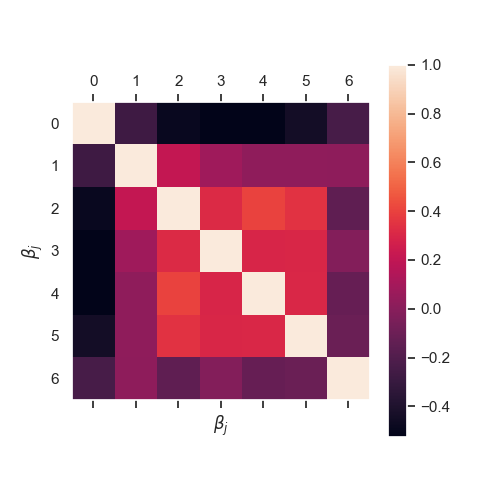}
    \caption{MCMC correlation matrix}\label{}
\end{subfigure}
\begin{subfigure}[t]{0.45\textwidth}
    \centering
    \includegraphics[width=\textwidth]{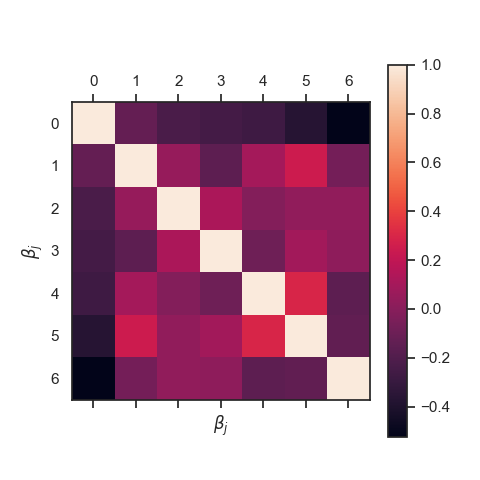}
    \caption{ABC correlation matrix}\label{}
\end{subfigure}
\caption{Correlation matrices for the heterogeneous rates simulation experiment.}
\label{fig:het-corr}
\end{figure}

\begin{table}
    \caption{Estimated marginal posterior standard deviations of heterogeneous infection rates (log-scale).}
    \centering
\begin{tabular}{lrrrr}
\toprule
Transmission Rate & Value & MCMC & NPE & ABC \\
\midrule
    Facility & -3.00 & 0.1941 & 0.178 & 0.264 \\
     Floor 1 & -3.91 & 0.711 & 0.808 & 0.773 \\
     Floor 2 & -3.22 & 0.449 & 0.724 & 0.896 \\
     Floor 3 & -2.81 & 0.756 & 0.761 & 0.865 \\
     Floor 4 & -2.53 & 0.474 & 0.654 & 0.973 \\
    Floor 5 & -2.30 & 0.259 & 0.462 & 0.844 \\
    Room & -3.00 & 0.278 & 0.658 & 1.01 \\
\bottomrule
\end{tabular}
\label{tab:het-sds}
\end{table}


\begin{table}
    \caption{Mean squared posterior predictive error for the heterogeneous rates simulation experiment.}
    \centering
    \begin{tabular}{lrrrr}
    \toprule
    Descriptive & Prior & MCMC & NPE & ABC \\
    \midrule
    Facility Incidence & 4330 & 267 & 482 & 389 \\
    Floor 1 Incidence & 146 & 24.9 & 31.6 & 38.8 \\
    Floor 2 Incidence & 195 & 47.5 & 53.4 & 51.4 \\
    Floor 3 Incidence & 190 & 32.7 & 47.7 & 46.7 \\
    Floor 4 Incidence & 223 & 40.3 & 49.4 & 53.2 \\
    Floor 5 Incidence & 347 & 54.8 & 76.8 & 79.1 \\
    Infected Rooms & 921 & 63.7 & 119 & 114 \\
    \bottomrule
    \end{tabular}
\end{table}

\subsection{CRKP Data Analysis}
\label{supp:crkp-figures}

\begin{table}
\caption{Estimates of the heterogeneous CRKP infection rates.}
\centering
\begin{tabular}{lrrrrr}
\toprule
Rate & MLE & Prior & MCMC & NPE & ABC \\
\midrule
Facility & 0.112 & 0.0821 & 0.112 & 0.0618 & 0.114 \\
Floor 1 & 0 & 0.0302 & 0.0144 & 0.0438 & 0.0367 \\
Floor 2 & 0.0657 & 0.0302 & 0.0328 & 0.0330 & 0.0311 \\
Floor 3 & 0 & 0.0302 & 0.0123 & 0.0997 & 0.0450 \\
Floor 4 & 0 & 0.0302 & 0.0147 & 0.0328 & 0.0286 \\
SCU & 0.0572 & 0.0302 & 0.0342 & 0.0363 & 0.0342 \\
Room & 0 & 0.0111 & 0.00605 & 0.0214 & 0.0125 \\
\bottomrule
\end{tabular}
\label{tab:crkp-rates}
\end{table}

\begin{figure}
\centering
\begin{subfigure}[t]{0.45\textwidth}
    \centering
    \includegraphics[width=\textwidth]{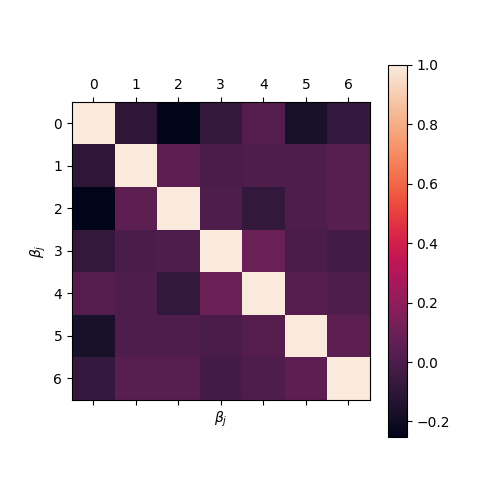}
    \caption{MCMC correlation matrix}\label{}
\end{subfigure}
\begin{subfigure}[t]{0.45\textwidth}
    \centering
    \includegraphics[width=\textwidth]{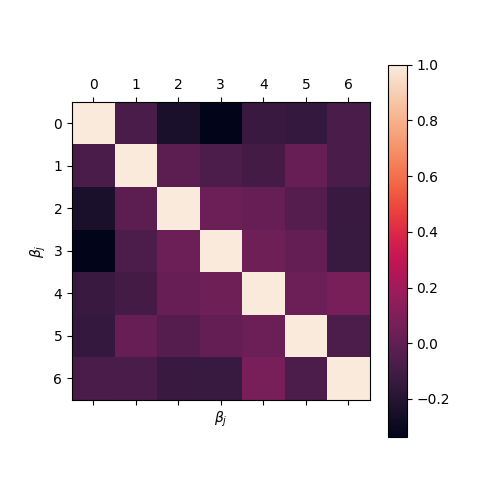}
    \caption{NPE correlation matrix}\label{}
\end{subfigure}
\caption{Correlation matrices for the heterogeneous CRKP transmission rates.}
\label{fig:crkp-corr}
\end{figure}

\begin{table}
\caption{Descriptive statistics for patients on LTACH Floors.}
\centering
\begin{tabular}{lcccc}
\toprule
Location & Admitted & Discharged & Imported Cases & Acquired Cases \\
\midrule
Floor 1 & 127 & 179 & 16 & 13 \\
Floor 2 & 334 & 256 & 48 & 51 \\
Floor 3 & 119 & 188 & 40 & 10 \\
Floor 4 & 176 & 165 & 29 & 16 \\
SCU & 134 & 102 & 20 & 16 \\
\bottomrule
\end{tabular}
\label{tab:ltach-floors}
\end{table}

\begin{table}
\caption{Flow of admitted patients between floors of the LTACH.}
\centering
\begin{tabular}{lcccc}
\toprule
Location & Patients Departing & Infecteds Departing & Patients Arriving & Infecteds Arriving \\
\midrule
Floor 1 & 98 & 33 & 150 & 76 \\
Floor 2 & 259 & 103 & 171 & 13 \\
Floor 3 & 133 & 66 & 216 & 147 \\
Floor 4 & 114 & 39 & 106 & 11 \\
SCU & 229 & 82 & 190 & 76 \\
\bottomrule
\end{tabular}
\label{tab:ltach-flow}
\end{table}

\FloatBarrier

\section{CRKP Transmission Predictive Checks}
\label{supp:crkp-checks}

\begin{figure}
\begin{subfigure}[t]{.4\textwidth}
\centering
\includegraphics[width=\linewidth]{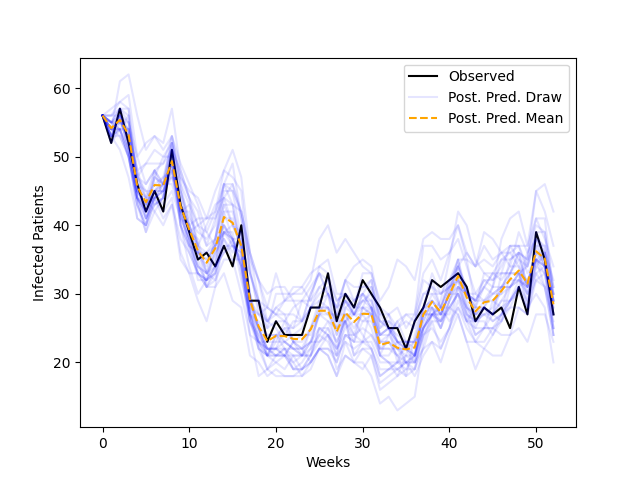}
\caption{Facility}
\label{}
\end{subfigure}
\hfill
\begin{subfigure}[t]{.4\textwidth}
\centering
\includegraphics[width=\linewidth]{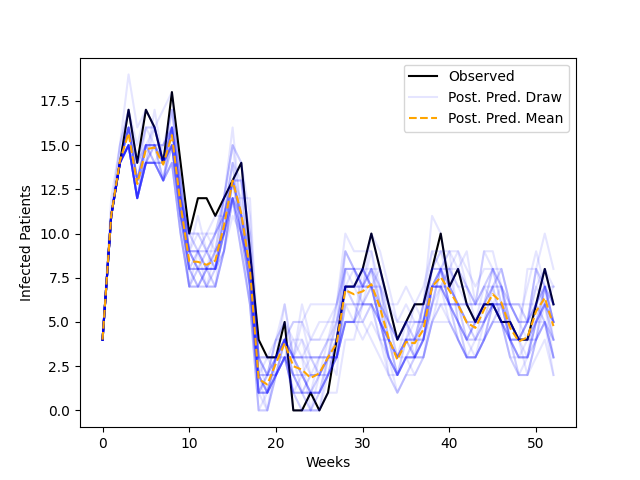}
\caption{Floor 1}
\label{}
\end{subfigure}

\medskip

\begin{subfigure}[t]{.4\textwidth}
\centering
\includegraphics[width=\linewidth]{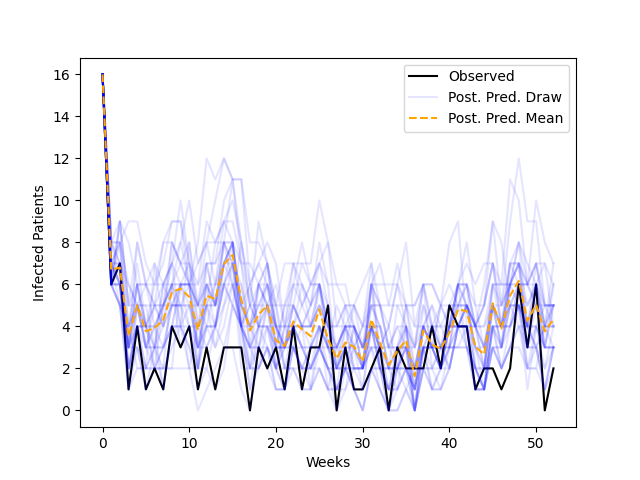}
\caption{Floor 2}
\label{}
\end{subfigure}
\hfill
\begin{subfigure}[t]{.4\textwidth}
\centering
\includegraphics[width=\linewidth]{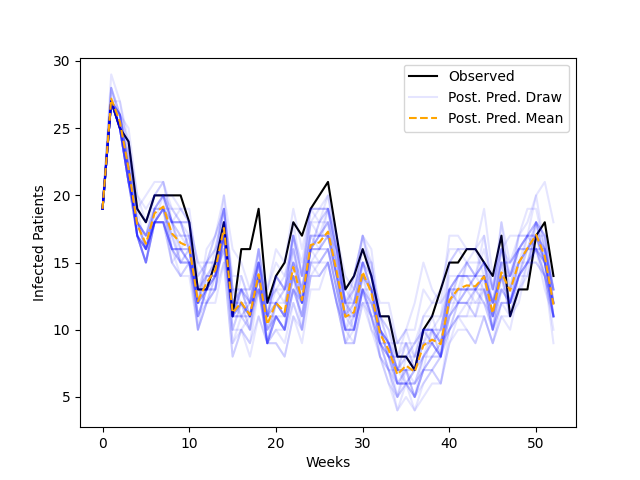}
\caption{Floor 3}
\label{}
\end{subfigure}

\medskip

\begin{subfigure}[t]{.4\textwidth}
\centering
\includegraphics[width=\linewidth]{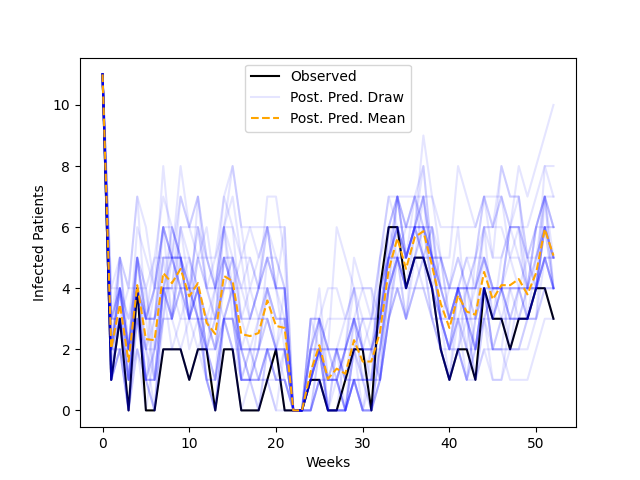}
\caption{Floor 4}
\label{}
\end{subfigure}
\hfill
\begin{subfigure}[t]{.4\textwidth}
\centering
\includegraphics[width=\linewidth]{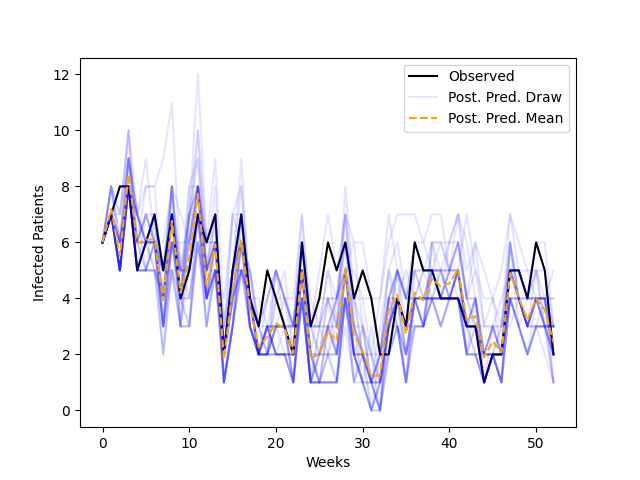}
\caption{SCU (Floor 5)}
\label{}
\end{subfigure}

\par

\centering
\begin{subfigure}[t]{.4\textwidth}
\centering
\includegraphics[width=\linewidth]{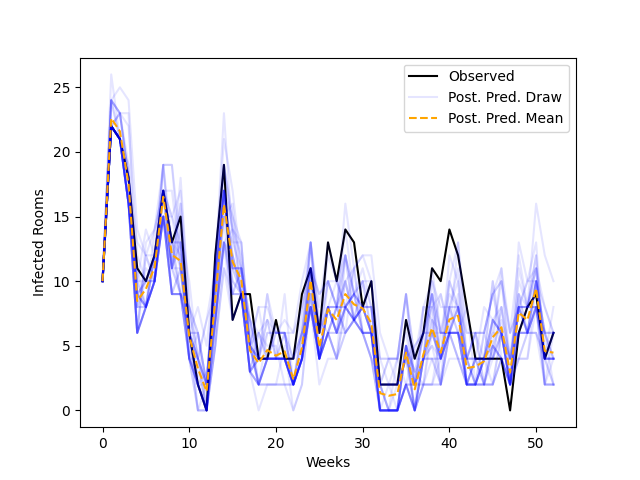}
\caption{Room}
\label{fig:crkp-ppc-homog-complete}
\end{subfigure}

\caption{Posterior predictive checks of the NPE-estimated homogeneous infection rate for the CRKP dataset.}
\end{figure}
\label{}

\begin{figure}
\begin{subfigure}[t]{.4\textwidth}
\centering
\includegraphics[width=\linewidth]{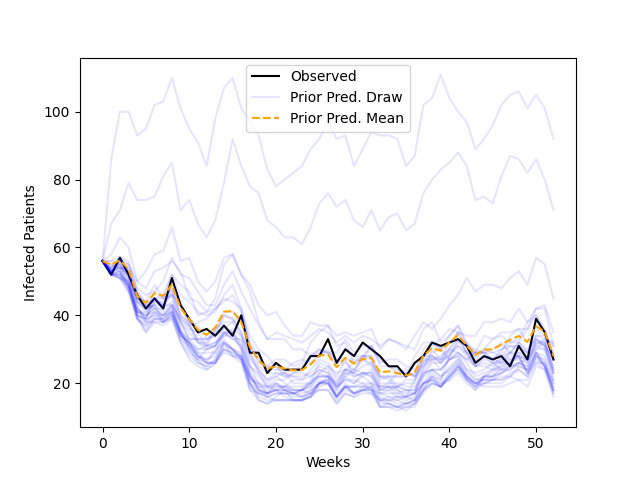}
\caption{Facility}
\label{}
\end{subfigure}
\hfill
\begin{subfigure}[t]{.4\textwidth}
\centering
\includegraphics[width=\linewidth]{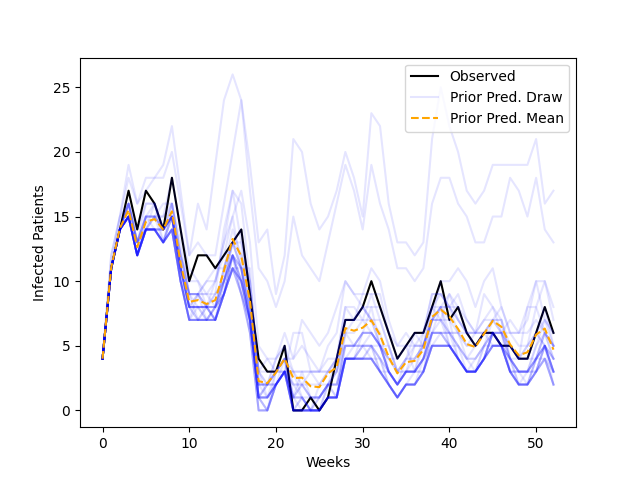}
\caption{Floor 1}
\label{}
\end{subfigure}

\medskip

\begin{subfigure}[t]{.4\textwidth}
\centering
\includegraphics[width=\linewidth]{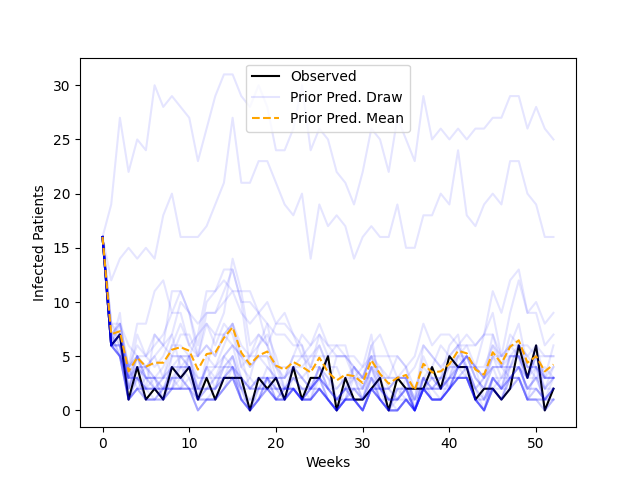}
\caption{Floor 2}
\label{}
\end{subfigure}
\hfill
\begin{subfigure}[t]{.4\textwidth}
\centering
\includegraphics[width=\linewidth]{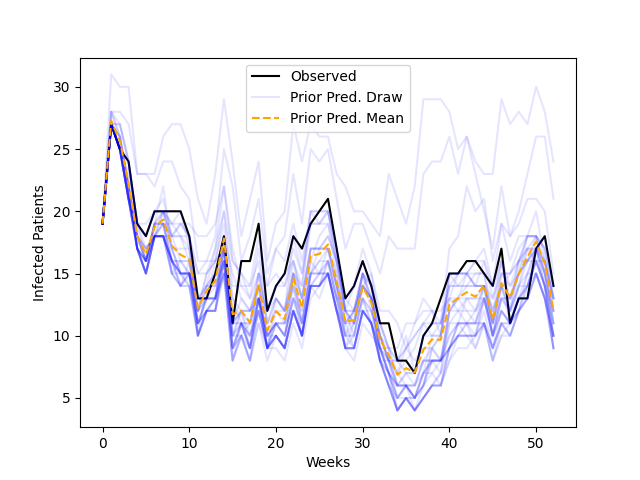}
\caption{Floor 3}
\label{}
\end{subfigure}

\medskip

\begin{subfigure}[t]{.4\textwidth}
\centering
\includegraphics[width=\linewidth]{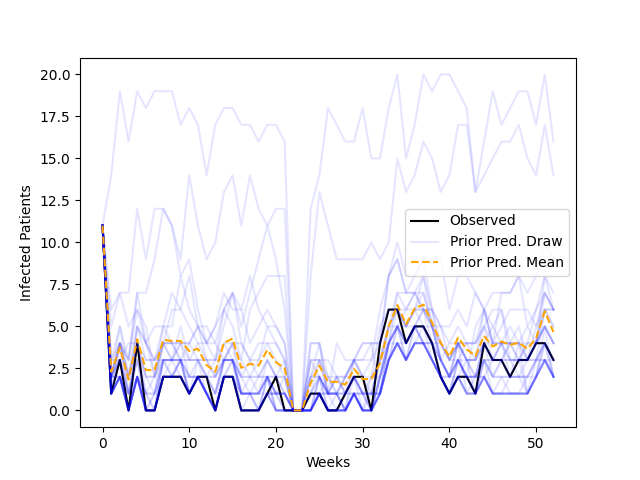}
\caption{Floor 4}
\label{}
\end{subfigure}
\hfill
\begin{subfigure}[t]{.4\textwidth}
\centering
\includegraphics[width=\linewidth]{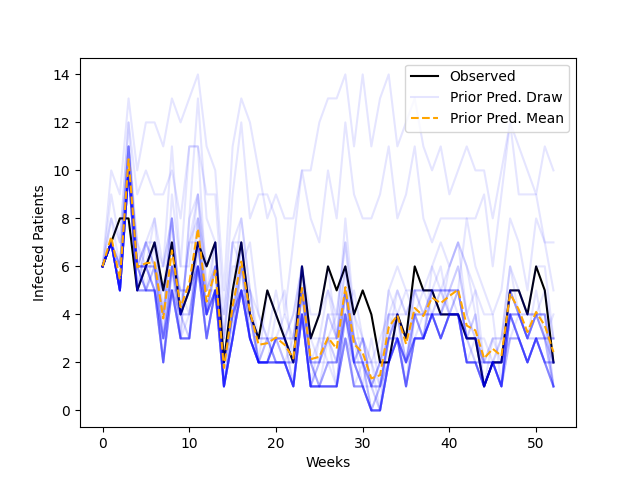}
\caption{SCU (Floor 5)}
\label{}
\end{subfigure}

\par

\centering
\begin{subfigure}[t]{.4\textwidth}
\centering
\includegraphics[width=\linewidth]{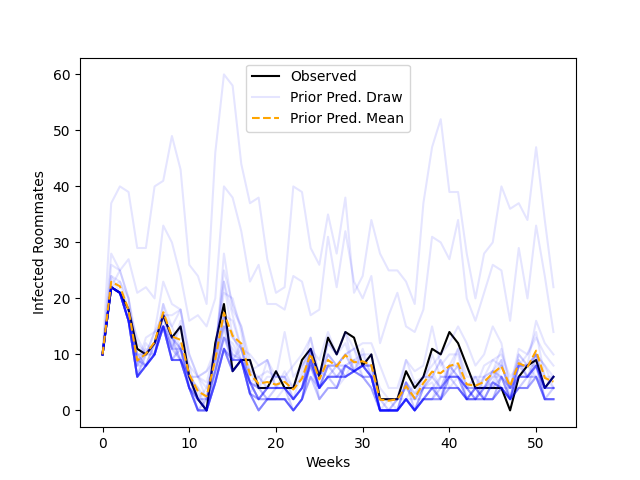}
\caption{Room}
\label{}
\end{subfigure}

\caption{Prior predictive checks of the heterogeneous infection rate for the CRKP dataset.}
\end{figure}
\label{}

\begin{figure}
\begin{subfigure}[t]{.4\textwidth}
\centering
\includegraphics[width=\linewidth]{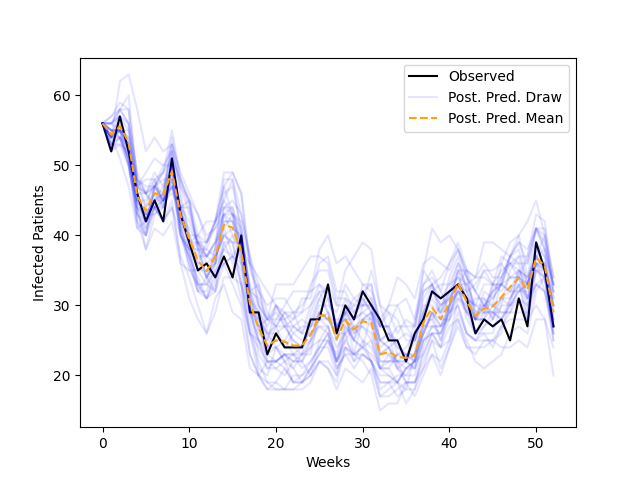}
\caption{Facility}
\label{}
\end{subfigure}
\hfill
\begin{subfigure}[t]{.4\textwidth}
\centering
\includegraphics[width=\linewidth]{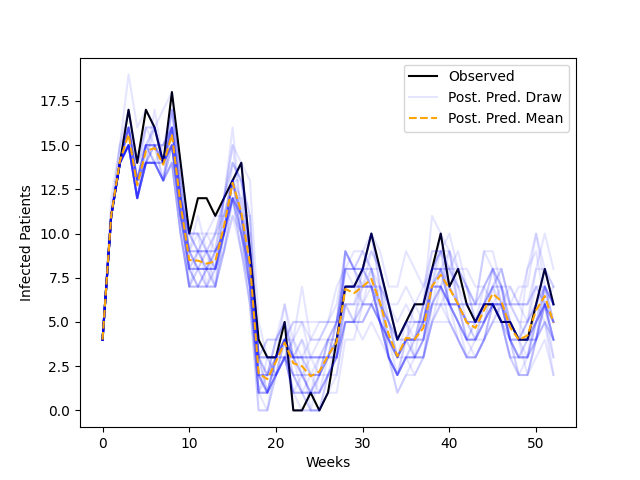}
\caption{Floor 1}
\label{}
\end{subfigure}

\medskip

\begin{subfigure}[t]{.4\textwidth}
\centering
\includegraphics[width=\linewidth]{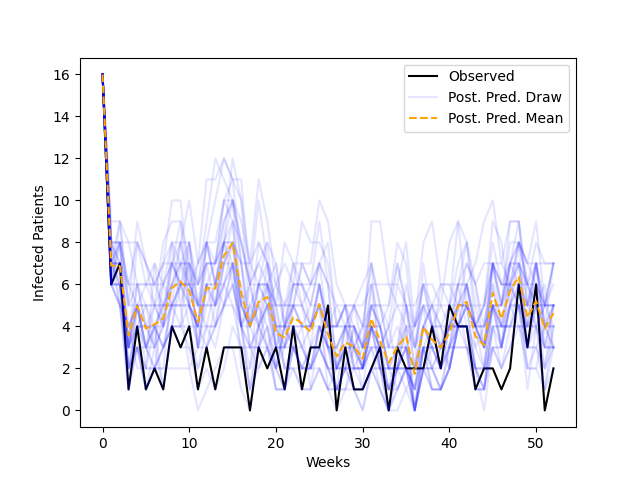}
\caption{Floor 2}
\label{}
\end{subfigure}
\hfill
\begin{subfigure}[t]{.4\textwidth}
\centering
\includegraphics[width=\linewidth]{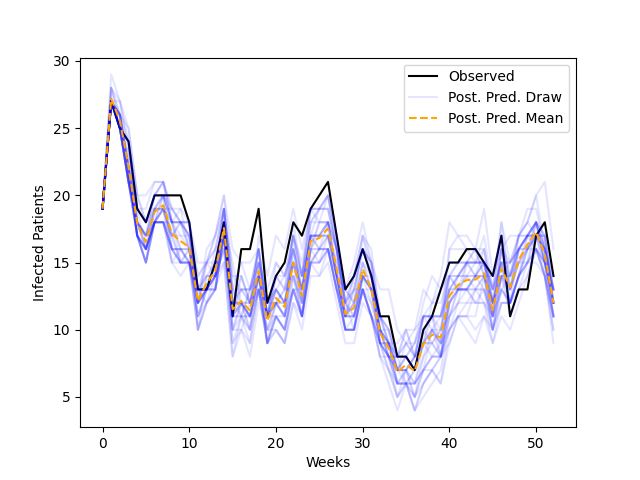}
\caption{Floor 3}
\label{}
\end{subfigure}

\medskip

\begin{subfigure}[t]{.4\textwidth}
\centering
\includegraphics[width=\linewidth]{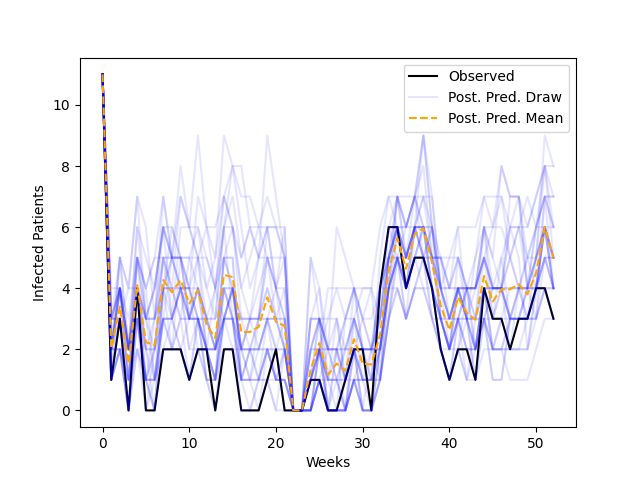}
\caption{Floor 4}
\label{}
\end{subfigure}
\hfill
\begin{subfigure}[t]{.4\textwidth}
\centering
\includegraphics[width=\linewidth]{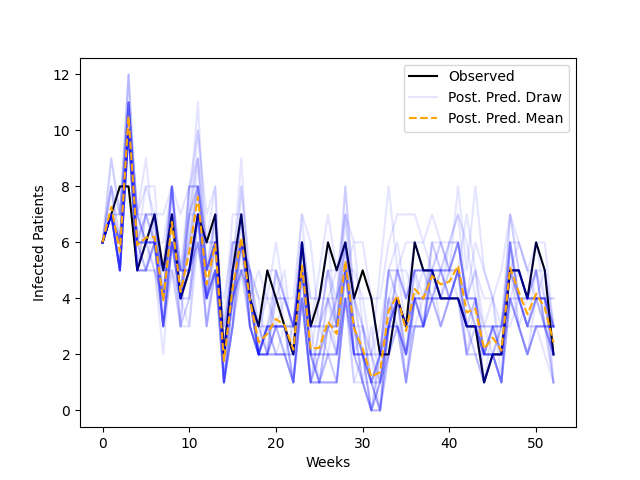}
\caption{SCU (Floor 5)}
\label{}
\end{subfigure}

\par

\centering
\begin{subfigure}[t]{.4\textwidth}
\centering
\includegraphics[width=\linewidth]{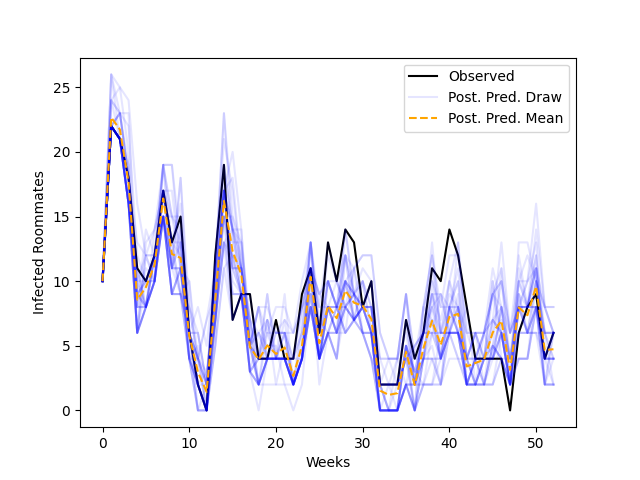}
\caption{Room}
\label{}
\end{subfigure}

\caption{Posterior predictive checks of the MCMC-estimated heterogeneous infection rates for the CRKP dataset.}
\end{figure}

\begin{figure}
\begin{subfigure}[t]{.4\textwidth}
\centering
\includegraphics[width=\linewidth]{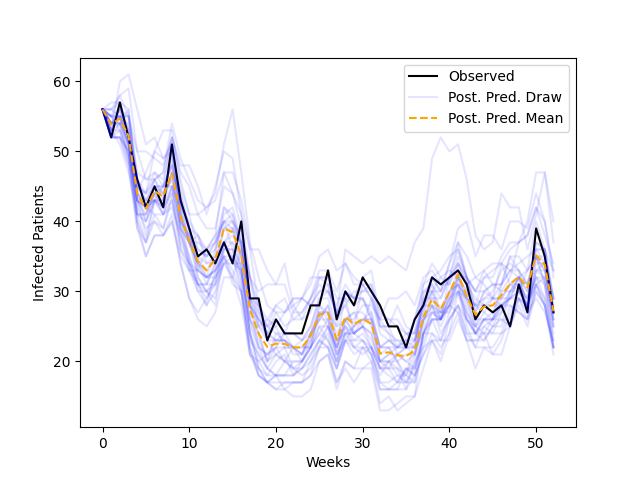}
\caption{Facility}
\label{}
\end{subfigure}
\hfill
\begin{subfigure}[t]{.4\textwidth}
\centering
\includegraphics[width=\linewidth]{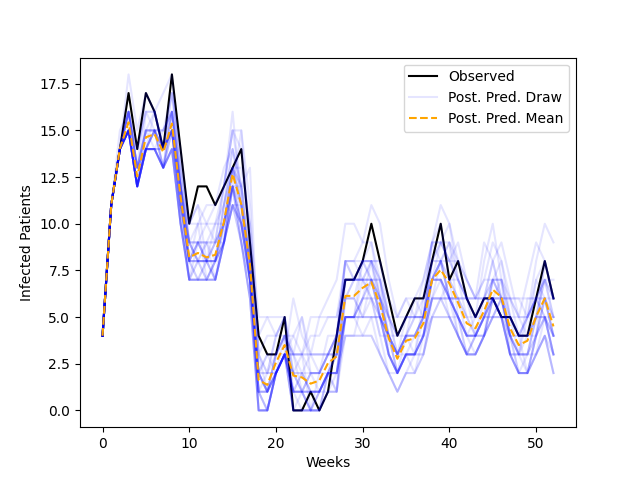}
\caption{Floor 1}
\label{}
\end{subfigure}

\medskip

\begin{subfigure}[t]{.4\textwidth}
\centering
\includegraphics[width=\linewidth]{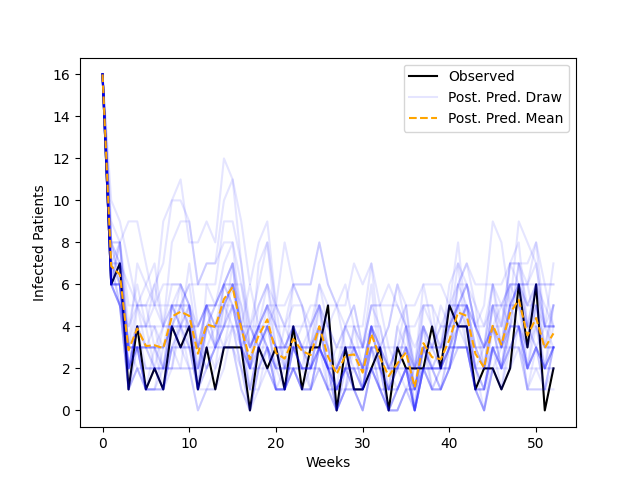}
\caption{Floor 2}
\label{}
\end{subfigure}
\hfill
\begin{subfigure}[t]{.4\textwidth}
\centering
\includegraphics[width=\linewidth]{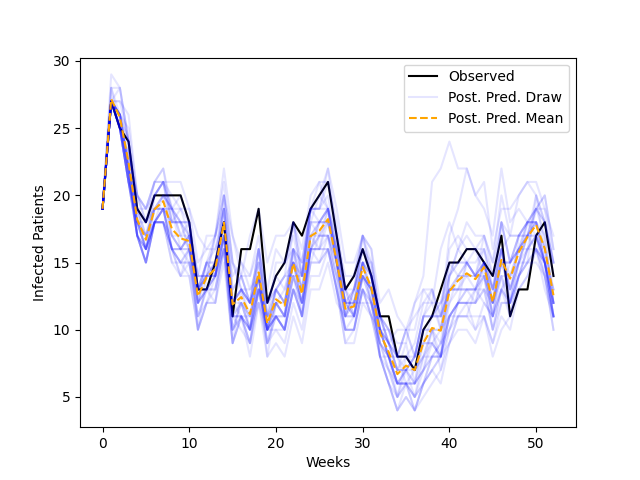}
\caption{Floor 3}
\label{}
\end{subfigure}

\medskip

\begin{subfigure}[t]{.4\textwidth}
\centering
\includegraphics[width=\linewidth]{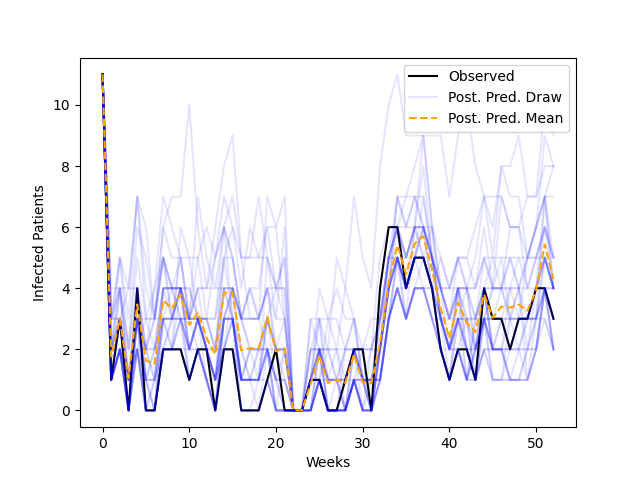}
\caption{Floor 4}
\label{}
\end{subfigure}
\hfill
\begin{subfigure}[t]{.4\textwidth}
\centering
\includegraphics[width=\linewidth]{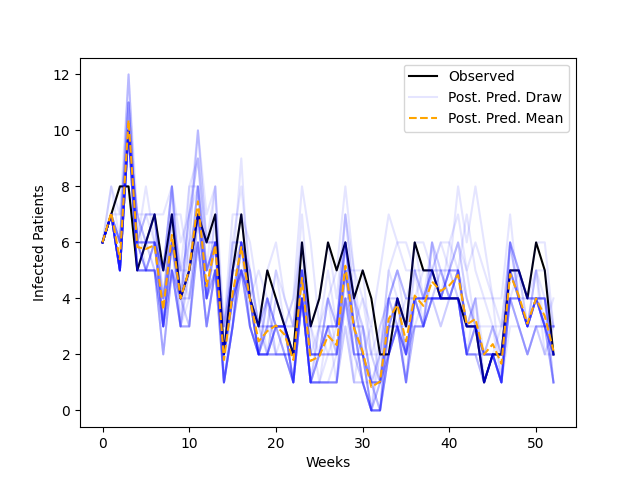}
\caption{SCU (Floor 5)}
\label{}
\end{subfigure}

\par

\centering
\begin{subfigure}[t]{.4\textwidth}
\centering
\includegraphics[width=\linewidth]{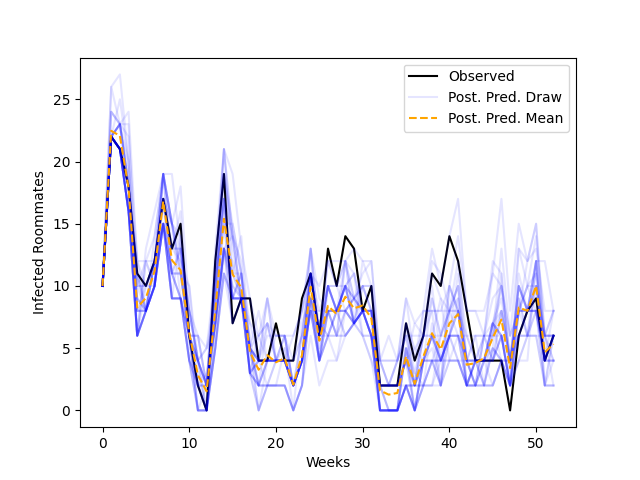}
\caption{Room}
\label{}
\end{subfigure}

\caption{Posterior predictive checks of the NPE-estimated heterogeneous infection rates for the CRKP dataset.}
\end{figure}

\begin{figure}
\begin{subfigure}[t]{.4\textwidth}
\centering
\includegraphics[width=\linewidth]{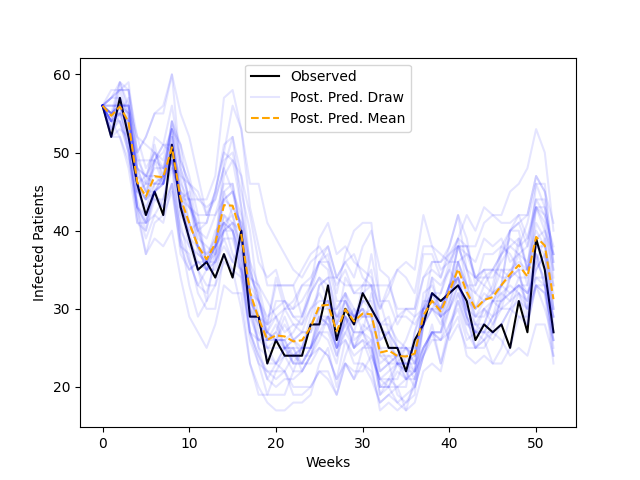}
\caption{Facility}
\label{}
\end{subfigure}
\hfill
\begin{subfigure}[t]{.4\textwidth}
\centering
\includegraphics[width=\linewidth]{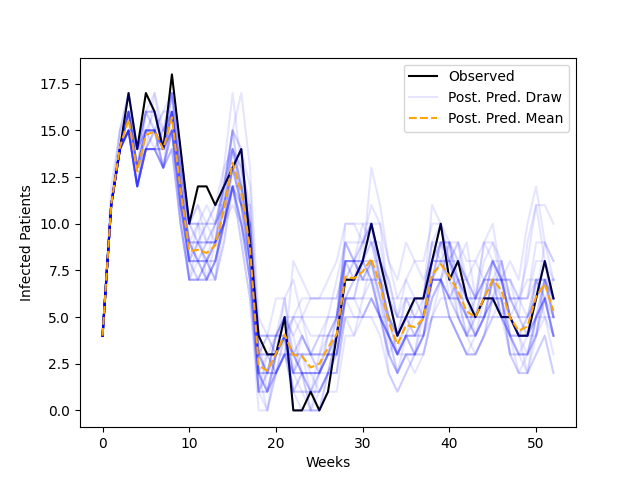}
\caption{Floor 1}
\label{}
\end{subfigure}

\medskip

\begin{subfigure}[t]{.4\textwidth}
\centering
\includegraphics[width=\linewidth]{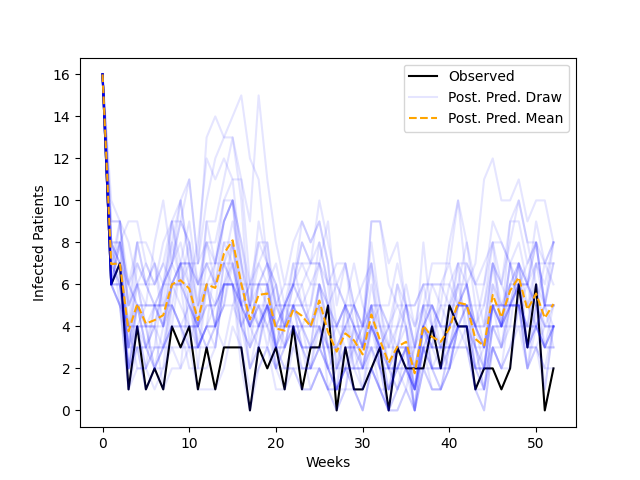}
\caption{Floor 2}
\label{}
\end{subfigure}
\hfill
\begin{subfigure}[t]{.4\textwidth}
\centering
\includegraphics[width=\linewidth]{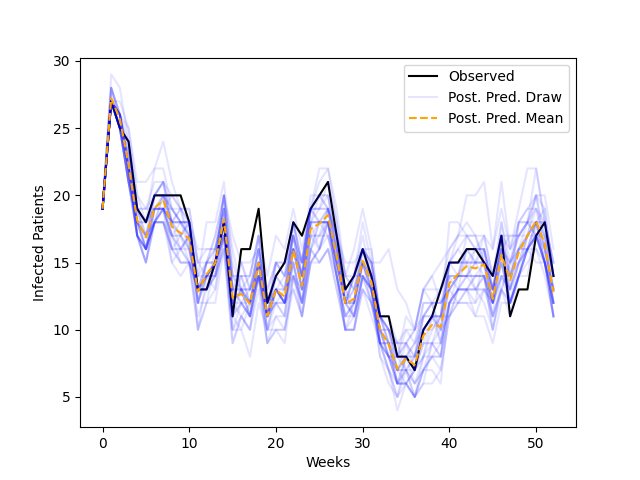}
\caption{Floor 3}
\label{}
\end{subfigure}

\medskip

\begin{subfigure}[t]{.4\textwidth}
\centering
\includegraphics[width=\linewidth]{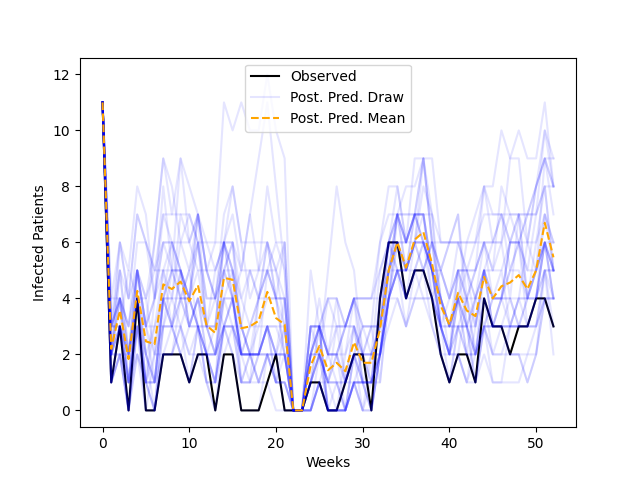}
\caption{Floor 4}
\label{}
\end{subfigure}
\hfill
\begin{subfigure}[t]{.4\textwidth}
\centering
\includegraphics[width=\linewidth]{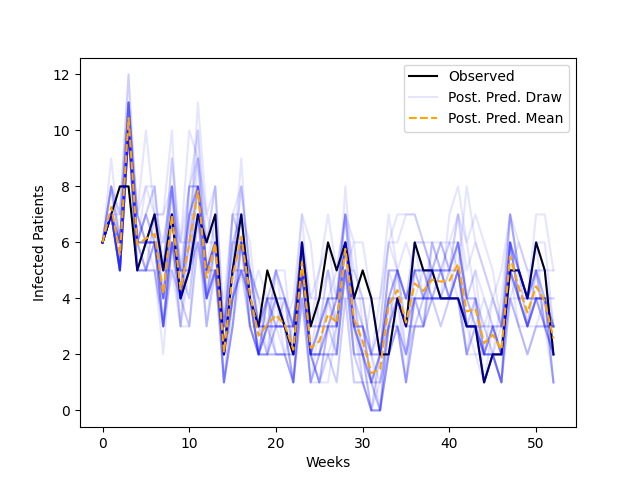}
\caption{SCU (Floor 5)}
\label{}
\end{subfigure}

\par

\centering
\begin{subfigure}[t]{.4\textwidth}
\centering
\includegraphics[width=\linewidth]{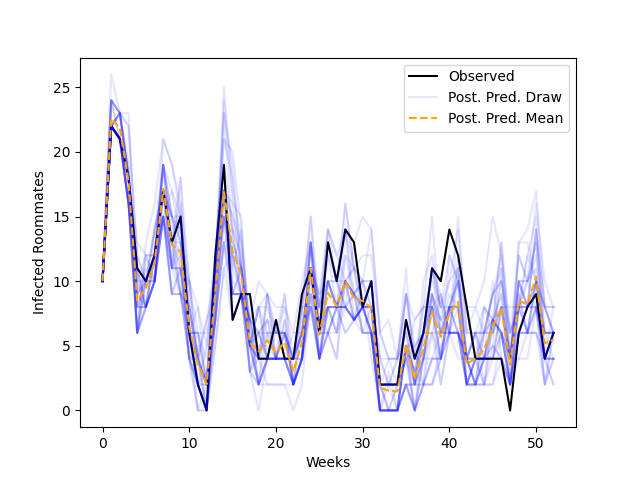}
\caption{Room}
\label{}
\end{subfigure}

\caption{Posterior predictive checks of the ABC-estimated heterogeneous infection rates for the CRKP dataset.}
\end{figure}

\end{appendix}

\end{document}